\documentclass[a4paper,11pt]{article} 
\usepackage{jheppub} 
\usepackage{comment} 
\usepackage{tikz} 
\usepackage{hyperref} 
\usepackage{mathrsfs}
\usepackage{mycommands} 
\usepackage{graphicx} 
\usepackage{mathtools}

\usepackage{color}

\title{Internal multiplicity distributions of jets from nonlinear evolution within the jet function framework}

\affiliation[a]{Key Laboratory of Quark and Lepton Physics (MOE) \& Institute of Particle Physics, Central China Normal University, Wuhan 430079, China}

\affiliation[b]{Key Laboratory of Particle Physics and Particle Irradiation (MOE),
Institute of frontier and interdisciplinary science,
Shandong University, Qingdao, Shandong 266237, China}

\author[a]{Pi Duan}
\emailAdd{duanpi@mails.ccnu.edu.cn}
\author[a]{Weiyao Ke}
\emailAdd{weiyaoke@ccnu.edu.cn}
\author[a]{Guang-You Qin}
\emailAdd{guangyou.qin@ccnu.edu.cn}
\author[a,  b]{Lei Wang}
\emailAdd{leiwangqd@sdu.edu.cn}

\abstract{
Jets selected with high internal charged-particle multiplicity exhibit markedly different substructure patterns compared to inclusive jet samples. Such correlations motivate a systematic study of jet observables as a function of the normalized multiplicity, $\nu = N_{\rm ch}/\langle N_{\rm ch}\rangle$.
In this work, we develop a theoretical framework for the full charged-particle multiplicity distribution of exclusive and inclusive jets, formulated within the jet-function approach. The hard production and jet function are evaluated at NLO + LL$_R$ accuracy. The internal parton dynamics governing the multiplicity distribution are described by coupled nonlinear branching equations with angular ordering, supplemented by a nonperturbative modeling term that accounts for hadron-level effects.
The resulting predictions are validated against \textsc{Pythia8} simulations and compared with CMS data. We examine the effects of both nonperturbative and perturbative components in shaping the multiplicity distribution, and show that Koba–Nielsen–Olesen (KNO) scaling is notably violated in the region $\nu > 2$ in the full solution, with a trend consistent with Monte Carlo results.
This framework that numerically solves the nonlinear multiplicity evolution goes beyond DLA-like approximations and reproduces key features seen in event generators, providing a solid foundation for future investigations of multiplicity-conditioned jet substructure within the jet function formalism.
}

\begin{document}
\maketitle
\flushbottom

\section{Introduction}
\label{sec:introduction}

Recent analyses by the CMS Collaboration have uncovered a fascinating class of jets characterized by exceptionally high internal charged-particle multiplicity in proton-proton ($p$-$p$) collisions~\cite{CMS:2023iam,CMS-PAS-HIN-24-024}. In these studies, jets with a cone size of $R=0.8$ and transverse momentum $p_{T, \rm jet}>550$~GeV exhibit large event-by-event fluctuations, with the normalized charged-particle multiplicity $\nu = N_{\rm ch}/\langle N_{\rm ch} \rangle$ reaching values up to five times the mean. These observations, motivated by the search for collective phenomena inside jets~\cite{Baty:2021ugw}, reveal distinctive features that challenge our current understanding of jet formation.

The internal jet multiplicity---defined as the number of charged tracks within the jet cone---should be distinguished from the global event multiplicity used in studies of long-range collectivity in high-multiplicity $p$-$p$ collisions~\cite{CMS:2013ycn,ALICE:2013tla,Azarkin:2014cja,Varga:2018isd,ALICE:2022jbp,ALICE:2023lyr,ALICE:2023oww}. Jets selected with high normalized multiplicity ($\nu \gtrsim 4$) display marked differences in their substructure compared to unselected or moderate-multiplicity jets. Notably, their charged-hadron fragmentation function develops a single-hump shape, deviating from the characteristic power-law behavior of inclusive jets~\cite{CMS:2023iam}. Furthermore, the two-particle correlation function $c_2\{2\}$ in the jet-centric frame shows a pronounced near-side ridge at large $\nu$, an effect not reproduced by standard Monte Carlo (MC) event generators like \textsc{Pythia8}~\cite{Bierlich:2022pfr} or \textsc{Sherpa}~\cite{Sherpa:2019gpd}. Recent substructure engineering analyses indicate that this enhanced ridge structure in $c_2\{2\}$ at large $\nu$ is associated with jets having a clear two-prong structure~\cite{CMS-PAS-HIN-24-024}.

Several theoretical interpretations have been proposed to explain these novel phenomena. One suggestion is that final-state interactions among jet constituents at both partonic and hadronic levels could generate such a ridge~\cite{Zhao:2024wqs}, drawing parallels with explanations for collective effects in small collision systems~\cite{Li:2012hc,Dusling:2015gta,Nagle:2018nvi,CMS:2010ifv,ATLAS:2015hzw,CMS:2015fgy,CMS:2016fnw}. An alternative proposal is that the high-multiplicity selection preferentially tags jets initiated by two hard partons~\cite{Dokshitzer:2025owq}. However, a more fundamental issue precedes these substructure studies: the measured multiplicity distribution itself reveals that the probability of observing jets with $\nu > 4$ significantly exceeds predictions from current Monte Carlo simulations and semi-analytic calculations~\cite{Dokshitzer:2025owq,Zhao:2024wqs,Baty:2021ugw}. This quantitative discrepancy highlights a critical shortfall in our ability to describe extreme multiplicity fluctuations.

The theoretical description of high-multiplicity jets faces profound challenges~\cite{KHOZE1997179,Dokshitzer:1982xr,Mueller:1982cq,Dokshitzer:1991wu,DOKSHITZER1993295,Dokshitzer:2005ri,Dokshitzer:2025fky,Duan:2025lvi,Duan:2025ngi}. Most analytical approaches rely on approximations such as the double-logarithmic (DLA), modified DLA, or modified leading logarithmic approximation (MLLA). Computationally, the extreme rarity of high-multiplicity jets makes direct MC generation prohibitively expensive; for jet energies around 500 GeV at 13 TeV collision energy, the probability for $\nu > 4$--$5$ is only $10^{-4}$--$10^{-5}$~\cite{Baty:2021ugw,Zhao:2024wqs}, rendering detailed studies of multi-particle correlations statistically challenging.

These dual challenges underscore the need for developing robust (semi-)analytic tools specifically focused on high-multiplicity jets. An ideal framework should: (1) accurately describe multiplicity distributions across the full kinematic range, from $\nu \approx 1$ to $\nu \gg 1$; and (2) enable systematic investigation of how jet substructures evolve with increasing multiplicity. This paper initiates a series of studies addressing these goals, focusing here on establishing the theoretical formulation for calculating jet multiplicity distributions and a discussion on the numerical results.

We employ the jet function approach --- a powerful method based on the Soft-Collinear Effective field Theory (SCET)~\cite{Bauer:2000ew,Bauer:2000yr,Bauer:2001ct,Bauer:2001yt,Beneke:2002ph}, which has been widely used to study jet cross sections and substructure observables~\cite{Becher:2008cf,Becher:2009cu,Becher:2009th,Ellis:2010rwa}. Our calculation combines next-to-leading order (NLO) fixed-order accuracy for the hard parton scattering cross sections~\cite{AVERSA1989105,Jager:2002xm} with leading-logarithmic (LL$_{R}$) resummation of jet cone size $R$~\cite{Dasgupta:2014yra,Dasgupta:2016bnd,Kang:2016mcy,Kang:2016ehg}. The multiplicity distribution inside the jet is obtained by solving a set of angular-ordered, coupled nonlinear evolution equations for the multiplicity generating functional, which goes beyond the usual double-logarithmic approximation. 
We then employ this framework to compare our calculations with recent CMS data and \textsc{Pythia8} simulations, and investigate the Koba–Nielsen–Olesen (KNO) scaling and its breaking~\cite{Koba:1972ng,Lam:1983vw,Hinz:1985wq,Hegyi:2000sp,H1:2020zpd,Vertesi:2020utz,Liu:2022bru,Liu:2023eve,Germano:2024ier} . Furthermore, we quantify the influence of different nonperturbative hadronization models on the predicted multiplicity distribution.

The remainder of this paper is structured as follows. Sec.~\ref{sec:generatingFunc} introduces the generating function formalism for jet multiplicity. In Sec.~\ref{sec:exclusiveJet}, we develop the theoretical framework for exclusive jets, detailing the calculation of NLO corrections, the derivation of renormalization group equations, and the implementation of angular-ordered evolution. The corresponding numerical results for exclusive jets and a study of key features like KNO scaling and flavor dependence are presented in Sec.~\ref{sec:resultsExclusive}. Sec.~\ref{sec:inclusiveJet} generalizes this framework to semi-inclusive jets. Our predictions are then compared with experimental data and \textsc{Pythia8} simulations in Sec.~\ref{sec:resultsSemiInclusive}. We finally summarize and conclude in Sec.~\ref{sec:conclusion}.

\section{Multiplicity distribution and generating function}\label{sec:generatingFunc}
Consider a parton of species ``$i$'' with energy $\omega$ and energy resolution scale $\mu$ that fragments into $n$ hadrons (or charged hadrons, and charged tracks) with probability $P_i(n, \omega, \mu)$. The generating function (or a partition function) for the multiplicity distribution $P_i(n, \omega, \mu)$ is defined as:
\begin{align}
Z_i(s, \omega, \mu) =& \mathscr{Z}\left\{P\right\}(s) = \sum_{n=0}^\infty e^{-ns} P_i(n, \omega, \mu), \label{eq:genFunc}
\end{align}
where $s$ is the conjugate variable to the multiplicity of charged hadrons $n$. The conservation of probability requires that under any condition
\begin{align}
\sum_{n=0}^\infty P(n)=1\,,\,\, Z(s=0)=1.
\end{align}
Information about mean, standard deviation, and higher-order cumulants ($c_m$) can be extracted from the generating function via
\begin{align}
  c_1 &= \langle n\rangle = -\left.\frac{\partial\ln Z}{\partial s}\right|_{s=0}\,,\,\, c_2 = \Big\langle {\left(n-\langle n\rangle\right)}^2\Big\rangle = \left.\frac{\partial^2\ln Z}{\partial s^2}\right|_{s=0}\, \nonumber\\
    c_m &= {(-1)}^m \left.\frac{\partial^m\ln Z}{\partial s^m}\right|_{s=0}\,. \label{eq:cumulants}
\end{align}
The original distribution can also be reconstructed from the generating function via an inverse transformation,
\begin{align}\label{eq:invertLaplace}
P(n) = \frac{1}{2\pi i} \int_{s_0 - i\pi}^{s_0 + i\pi} Z(s) e^{ns} \mathrm{d} s \, ,
\end{align}
where $s_0$ lies to the right of any poles of $Z(s)$.

The advantage of studying $Z(s)$ is that convolutions in the particle number become simple products in $s$-space,
\begin{align}\label{eq:convolveToprodcut}
P_1\otimes_n P_2 \equiv& \sum_{k=0}^n P_1(k) P_2(n-k) \, ,\nonumber\\
\mathscr{Z}\left\{P_1\otimes_n P_2\right\}(s) =& \mathscr{Z}\left\{P_1\right\}(s) \, 
\times \, 
\mathscr{Z}\left\{P_2\right\}(s) \,.
\end{align}
This property is especially useful for implementing multiplicity constraints in jet-like systems with multiple independent fragmenting partons.

\paragraph{Operator definition of $Z_i$ for exclusive jet}
Even though the multiplicity distribution and generating function are clearly nonperturbative objects, and will certainty require some degree of modeling in this study, here we still provide an operator definition for them for completeness. The definition is presented within the framework of SCET~\cite{Bauer:2000ew,Bauer:2000yr,Bauer:2001ct,Bauer:2001yt}, with the basic building blocks of gauge-invariant collinear quark and gluon fields,
\begin{align}
  \chi_{n} = W_{n}^{\dagger} \xi_n, \quad  \mathcal{B}_{n \perp}^{\mu} = \frac{1}{g} {\left[ W_{n}^{\dagger} i D_{n \perp}^{\mu} W_n \right]}, \label{eq:SCETfield}
\end{align}
Here, $n^\mu=(1,0,0,1)$ is the light-cone vector whose spatial component aligns with the jet direction. The conjugate vector $\bar{n}^{\mu}=(1,0,0,-1)$, satisfying $n^2 = \bar{n}^2 = 0$, and $n \cdot \bar{n} = 2$. And $i D_{n \perp}^{\mu} = \mathcal{P}_{n\perp}^{\mu} + g A_{n \perp}^{\mu}$ is the transverse covariant derivative, where $\mathcal{P}^{\mu}$ is the label momentum operator. The Wilson line $W_n$, which ensures gauge invariance under collinear gauge transformations, is given by
\begin{align}
  W_n(x) = \sum_{\text{perms}} \exp {\left[ -g \frac{1}{\bar{n} \cdot \mathcal{P}} \bar{n} \cdot A_n(x)  \right]}.
\end{align}

We can now obtain exclusive jet multiplicity distributions and their generating functions for both quark and gluon jets. Here, exclusive means that all energy of the parent parton is contained within the cone $R$
\begin{align}\label{eq:exclusiveJetMultiDefinition}
  M_q(n, \omega, R) 
  &= \frac{1}{2d_F} \sum_{X_J} \delta(n-N_{J}^{\rm POI}) \Tr {\left[ \frac{\slashed{\bar{n}}}{2} \langle 0 | \delta(\omega - \bar{n} \cdot \mathcal{P}) \chi_n(0) |X_J\rangle \langle X_J| \bar{\chi}_n(0) | 0 \rangle \right]} \nonumber \\
  &\equiv J_q(\omega,R)P_q(n, \omega,R)\,, \\
  M_g(n, \omega, R) 
  &= -\frac{\omega}{(d-2)d_A} \sum_{X_J}\delta(n-N_{J}^{\rm POI})\langle 0 |\delta(\omega - \bar{n} \cdot \mathcal{P}) \mathcal{B}_{n\perp \mu}(0) |X_J \rangle \langle X_J| \mathcal{B}_{n\perp}^{\mu}(0) | 0 \rangle\nonumber\\
  &\equiv  J_g(\omega,R)P_g(n, \omega,R)\,, 
\end{align}
where $(d-2)$ is the number of physical gluon polarizations in $d$ space-time dimensions, $d_F=N_c$ and $d_A = N_c^2-1$. We restrict ourselves to massless quark flavors. The variable $\omega$ denotes the large light-cone momentum of the initiating parton $(\omega = \bar{n} \cdot p)$. $X_J$ labels the final state that satisfies the requirement for exclusive jet.
$\delta(n-N_{J}^{\rm POI})$ counts the number of particles of interest (POI) within the jet. For example, if the POI are charged particles, we will label it as $N_{J}^{\rm ch}$. We require all particles of interest to be within the jet cone, even for exclusive jet. This is because there is a maximum cut in the allowed energy leaks out of the cone for exclusive jet, and it is possible that $X_J$ satisfies the exclusive jet requirement, yet still contains soft hadrons out of the cone, which should not be counted in the internal multiplicity of the jet.

In the second line of each equation in Eq.~(\ref{eq:exclusiveJetMultiDefinition}), we factorize the result into the usual exclusive jet function $J_{i}(\omega, R)$ times the multiplicity probability distribution function $P_{i}(n,\omega, R)$.
The probability distribution is normalized so that $\sum_n P_i(n, \omega,R) = 1$, where $i$ represents a quark ($q$) or gluon ($g$) jet, respectively. 
Therefore, summing over $n$, the above definition will result in the exclusive jet function without conditioning on the multiplicity~\cite{Ellis:2010rwa}, i.e., $\sum_n M_{q,g} = J_{q,g}$. Following these, naturally, the operator definition of the generating function is
\begin{align}
\label{eq:formal-def-Exclusive-q}
  \tilde{M}_q(s, \omega, R) 
  &= \frac{1}{2d_F} \Tr {\left[ \frac{\slashed{n}}{2} \sum_{X_J} e^{-s N_J^{\rm ch}} \langle 0 | \delta(\omega - \bar{n} \cdot \mathcal{P}) \chi_n(0) |X_J \rangle \langle X_J| \bar{\chi}_n(0) | 0 \rangle \right]} \nonumber \\
  &\equiv J_q(\omega,R)Z_q(s, \omega,R) \, ,\\
  \tilde{M}_g(s, \omega, R) 
  &= -\frac{\omega}{(d-2)d_A} \sum_{X_J} e^{-s N_J^{\rm ch}}\langle 0 |\delta(\omega - \bar{n} \cdot \mathcal{P}) \mathcal{B}_{n\perp \mu}(0) |X_J \rangle \langle X_J| \mathcal{B}_{n\perp}^{\mu}(0) | 0 \rangle \nonumber\\
  &\equiv J_g(\omega,R)Z_g(s, \omega,R) \,.
\end{align}

\paragraph{Generating function for semi-inclusive jet} The definition of a semi-inclusive jet allows a measurable fraction $(1-z)$ of the energy of the original parton to leak outside the jet cone.
Thus, the generating function for the semi-inclusive jet is defined as
\begin{align}
  \tilde{M}_q(z,s,\omega_J,R) 
  &= \frac{z}{2d_F} \Tr {\left[ \frac{\slashed{\bar{n}}}{2} \sum_{X_J'} e^{-s N_J^{\rm ch}} \langle 0 | \delta(\omega_J/z - \bar{n} \cdot \mathcal{P}) \chi_n(0) |X_J' \rangle \langle X_J'| \bar{\chi}_n(0) | 0 \rangle \right]} \nonumber\\
  & \equiv J_q(z,\omega_J, R) Z_q(s,\omega_J,R) \, ,\\
   \tilde{M}_g(z,s,\omega_J, R) 
  &= -\frac{z\omega}{(d-2)d_A}\sum_{X_J'} e^{-s N_J^{\rm ch}}\langle 0 |\delta(\omega_J/z - \bar{n} \cdot \mathcal{P})  \mathcal{B}_{n\perp \mu}(0) |X_J' \rangle \langle X_J'| \mathcal{B}_{n\perp}^{\mu}(0) | 0 \rangle \nonumber\\
  &\equiv J_g(z,\omega_J, R) Z_g(s,\omega_J,R) \,.
\end{align}
Now, $X_J'$ denotes the state where a jet of cone size $R$ and energy $\omega_J$ can be reconstructed, but now allows any energy flow out of the cone. $\omega_J$ is the energy of the jet, carrying a fraction $z$ of the original parton ($\omega=\omega_J/z$). Again, the normal semi-inclusive jet function can be obtained by taking $s=0$~\cite{Kang:2016mcy,Kang:2016ehg}. The semi-inclusive jet multiplicity function depends on the jet radius $R$, but for notational simplicity, we will not explicitly write out the $R$-dependence in the remainder of this paper.

\paragraph{The nonperturbative nature of multiplicity distribution}
Multiplicity measures the number of charged hadrons, which are not fundamental degrees of freedom within SCET. Therefore, for practical calculations, it is necessary to further factorize the matrix element into two parts: one describing the partonic final state and the other accounting for the fragmentation of multi-parton states into hadrons with the specified multiplicity. For example, for the exclusive quark jet, the definition in Eq.~\eqref{eq:formal-def-Exclusive-q} is separated into
\begin{align}
& \tilde{M}_q(s, \omega, R)\nonumber\\
&= \sum_{X_J}\sum_{Y_J}e^{-s N_J^{\rm POI}} |\langle Y_J |X_J\rangle|^2 
\frac{1}{2d_F} \Tr \left\{ \frac{\slashed{n}}{2}\langle 0|\delta(\omega - \bar{n}\cdot\mathcal{P})\chi_n(0)|Y_J\rangle \langle Y_J|\bar{\chi}_n(0)|0\rangle\right\}\,,
\end{align}
where $Y_J$ denotes the partonic-level final state composed of quarks and gluons at a sufficiently high scale such that perturbative calculations remain valid. The factor $\langle X_J |Y_J\rangle$ represents the transition amplitude from the partonic to the hadronic state and encapsulates all nonperturbative information. Interference terms like $\langle X_J|Y_J'\rangle\langle Y_J|X_J\rangle$ have been neglected under the assumption that the multiplicity measurement is sufficiently inclusive to wash out phase correlations between distinct partonic configurations.

We further introduce an approximation regarding hadronization: hadron production is assumed to occur independently for each parton. Under this assumption, the matrix element can be simplified as
\begin{align}
 \tilde{M}_q(s, \omega, R) 
&= \frac{1}{2d_F} \sum_{Y_J} \Tr \left\{ \frac{\slashed{n}}{2} \langle 0|\delta(\omega - \bar{n}\cdot\mathcal{P})\chi_n(0)|Y_J\rangle \langle Y_J|\bar{\chi}_n(0)|0\rangle\right\}\prod_{j\in Y_J} Z_j(s)\,, \\
Z_j(s) &= \sum_{X_j} e^{-s N_j^{\rm POI}} |\langle X_j |j\rangle|^2\,.
\end{align}
This expression reduces the calculation to the familiar “jet calculus” framework~\cite{KONISHI197945,Kalinowski:1980ju}. For large charged-particle multiplicities and for fragmentation processes occurring at scales well above the hadronic scale, $\Lambda_{\rm QCD}$, the production of hadrons from different partons can be treated as statistically independent to leading order, since long-range color and kinematic correlations are power suppressed.
However, this approximation clearly fails in the absence of such scale separation: for instance, the two valence quarks inside a pion at nonperturbative scales are strongly correlated and must jointly form a single hadron---the pion itself.
Another situation where this assumption may break down is in highly populated phase spaces, such as events with extreme multiplicity, or with medium effects, where parton recombination processes may become relevant.
A detailed discussion of when and how parton recombination significantly affects particle production in high-multiplicity jets lies beyond the scope of this work.

\paragraph{Boundary condition of $Z(s)$}
The generating function $Z(s, \omega, \mu)$ is constrained by fundamental principles that dictate its boundary behavior. First, probability conservation mandates $Z(s=0, \omega, \mu)=1$. Second, a more subtle constraint arises from the existence of a physical mass gap: a parton with vanishing energy $(\omega \to 0)$ cannot produce massive hadrons. This necessitates that the multiplicity distribution reduces to a delta function at zero,
\begin{align}
P_i(n, \omega=0, \mu)=\delta_{n0}, \quad \text{or equivalently} \quad Z_i(s, \omega=0, \mu) = 1. \label{Eq.noEnergy}
\end{align}
Imposing this condition on the low scale $\mu \sim \Lambda_{\text{QCD}}$ highlights the inherent soft sensitivity of the multiplicity observable. A purely perturbative treatment, based on a power expansion in $\Lambda_{\rm QCD}/\omega$, would violate this physical boundary and exhibit soft divergences. Such soft sensitivity and associated logarithmic enhancement should be systematically resummed via angular ordering, as detailed in Sec.~\ref{sec:angularOrdering}. The primary motivation for adopting this physically-grounded boundary condition is to facilitate a robust exploration of nonperturbative models for $Z_q, Z_g$ and to quantitatively assess their phenomenological implications, which we will pursue in Sec.~\ref{sec:NPModel}.

\section{Internal multiplicity distribution of exclusive jets}\label{sec:exclusiveJet}
The multiplicity distribution of an exclusive jet involves two characteristic scales. The first is an initial infrared scale $Q_0 \gtrsim \Lambda_{\rm QCD}$, where the functional form of $Z(s,\omega, Q_0)$ must be determined from hadron-level physics. The second is the jet scale of order $p_T R$.
Throughout this work, we parametrize the initial condition in the regime $Q_0 \ll p_T R$. Performing a systematic power expansion in $Q_0/(p_T R)$, we treat all final-state partons as massless. Multiple emissions within the kinematic region between $Q_0$ and $p_T R$ will lead to a non-linear evolution equation for the generating function.

To describe collinear parton branchings in $d = 4 - 2\epsilon$ dimensions, we employ the spin-averaged $1 \to 2$ real-emission components of the Altarelli-Parisi splitting kernels~\cite{Altarelli:1977zs,Catani:1996vz}. In dimensional regularization, these are
\begin{subequations}\label{eq:DRAPsplitting}
\begin{align}
\hat{p}_{qq}(z,\epsilon) &= C_F \left[\frac{1+z^2}{1-z}-\epsilon(1-z) \right], \\
\hat{p}_{gq}(z,\epsilon) &= C_F \left[\frac{1+{(1-z)}^2}{z} -\epsilon z \right], \\
\hat{p}_{qg}(z,\epsilon) &= T_F \left[1 - \frac{2z(1-z)}{1-\epsilon} \right], \\
\hat{p}_{gg}(z,\epsilon) &= 2C_A \left[\frac{z}{1-z}+\frac{1-z}{z}+z(1-z)\right],
\end{align}
\end{subequations}
where $C_F = (N_c^2-1)/(2N_c)$, $C_A = N_c$, and $T_F = 1/2$. We also define the Altarelli-Parisi kernels that include the virtual contributions in the $\epsilon\rightarrow 0$ limit as
\begin{subequations}\label{eq:standAPsplitting}
\begin{align}
  \hat{P}_{qq}(z) &= C_F {\left[ \frac{1+z^2}{{(1-z)}_+} + \frac{3}{2} \delta (1-z) \right]}, \\
  \hat{P}_{gq}(z) &= C_F \frac{1 + {(1-z)}^2}{z}, \\
  \hat{P}_{qg}(z) &= T_F {\left[ z^2 + {(1-z)}^2 \right]}, \\
  \hat{P}_{gg}(z) &= C_A {\left[ \frac{2z}{{(1-z)}_+} + \frac{2(1-z)}{z} + 2z(1-z) \right]} + \frac{\beta_0^{n_f}}{2} \delta(1-z),
\end{align} 
\end{subequations}
where $\beta_0^{n_f} = \frac{11}{3}C_A - \frac{4}{3}n_f T_F$ is the QCD one-loop beta-function coefficient. The ``plus''-prescription is defined by
\begin{align}
  \int_{0}^{1} \mathrm{d} z f(z) {\left[ g(z) \right]}_+ \equiv \int_{0}^{1} \mathrm{d} z {\left[ f(z) - f(1) \right]}g(z)\,.
\end{align}
In what follows, we will express our final results in terms of these standard splitting functions in Eqs.~\eqref{eq:standAPsplitting}.

\begin{figure}
\centering
\includegraphics[width=.8\textwidth]{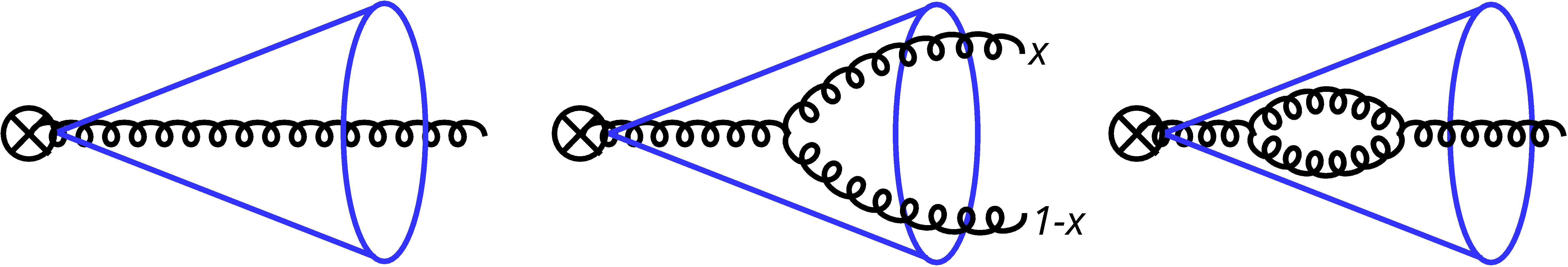}
\caption{Left: the LO diagram for $Z_g(s,\omega, \mu)$. Middle: the real emission amplitude. Its amplitude square contribute to the NLO correction $Z_g(s,\omega, \mu)$.
Right: the loop diagram that interfere with the LO diagram, which is scaleless under dimensional regularization. Diagrams with collinear emission from the Wilson line is not present for light-cone gauge.}\label{fig:g2gg-exclusive}
\end{figure}

\subsection{LO and NLO calculation in the pure-gluon system}
With these tools, we can compute the multiplicity distribution function for exclusive jets at NLO, taking the pure-gluon system as an example to demonstrate the key steps. We will use the axial gauge with $\bar{n}\cdot A=0$, this way, there are only three diagrams that contribute to LO and NLO and are shown in Fig.~\ref{fig:g2gg-exclusive}. The loop diagram is scaleless under dimensional regularization. 

The LO result is just a single gluon within the jet cone
\begin{align}
  \tilde{M}^{g,(0)}_g(s,\omega_J) = Z_g^{(0)}(s,\omega_J)
\end{align}
where $\omega=\omega_J$ for the exclusive jet. With radiative correction, the result can depend on the jet algorithm. We focus here on the anti-$k_T$ algorithm~\cite{Cacciari:2008gp}. In the case where both partons are inside the jet, the constraints imposed by the anti-$k_T$ algorithm with jet radius $R$ (with implicit $R$ dependence) can be expressed in terms of the following theta function~\cite{Kang:2016mcy, Ellis:2010rwa}
\begin{align}
\Theta_{\text{alg}}^{< R}
= \theta {\left( x(1-x) \omega_J \tan \frac{\mathcal{R}}{2} - k_{\perp} \right)} \label{eq:antiIn},
\end{align}
where $k_\perp$ is the transverse momentum of one of the daughter partons relative to the original parton, and $\mathcal{R}$ is related to the jet radius $R$
\begin{align}
\mathcal{R} \equiv \frac{R}{\cosh \eta}, \label{eq:jetR}
\end{align}
with the jet rapidity $\eta$. Then, the NLO contribution is
\begin{align}
  {M}_{g,\text{bare}}^{gg,(1)}(n, \omega_J) &= \frac{\alpha_s}{2\pi^2} \frac{{(\mu^2 e^{\gamma_E})}^{\epsilon}}{{(4\pi)}^{\epsilon}} \int_{0}^{1} \mathrm{d} x \frac{1}{2} \hat{p}_{gg}(x, \epsilon) \int \frac{\mathrm{d}^{2-2\epsilon} k_{\perp}}{{(2\pi)}^{-2\epsilon}}\frac{1}{k_{\perp}^{2}} \Theta_{\text{alg}}^{< R} \nonumber\\
& \quad \times \sum_{n_1=0}^{n} \sum_{n_2=0}^{n} P_g^{gg,(0)}(n_1, x\omega_J; n_2, (1-x)\omega_J) \delta_{n_1+n_2,n},
\end{align}
Here, the factor of $1/2$ accounts for the identical nature of the two final-state gluons. Assuming independent fragmentation, as justified above, the joint multiplicity distribution for the two daughter gluons factorizes into the product of their individual distributions:
\begin{align}
  P_g^{gg,(0)}(n_1, x\omega_J; n_2, (1-x)\omega_J) \approx P_{g}^{(0)}(n_1, x\omega_J) P_{g}^{(0)}( n_2, (1-x)\omega_J), \label{eq:indenp}
\end{align}
so that both final-state partons contribute to the total multiplicity, which introduces non-linearity in $Z$. After transforming into the generating function space as Eq.~\eqref{eq:genFunc} shows and using the property in Eq.~\eqref{eq:convolveToprodcut}, we get the NLO contribution of the exclusive jet multiplicity generating function
\begin{align}\label{eq:expressNLOpureg}
  \tilde{M}_{g,\text{bare}}^{gg,(1)}(s, \omega_J) &= \frac{\alpha_s}{2\pi^2} \frac{{(\mu^2 e^{\gamma_E})}^{\epsilon}}{{(4\pi)}^{\epsilon}} \int_{0}^{1} \mathrm{d} x \frac{1}{2} \hat{p}_{gg}(x, \epsilon) \int \frac{\mathrm{d}^{2-2\epsilon} k_{\perp}}{{(2\pi)}^{-2\epsilon}}\frac{1}{k_{\perp}^{2}} \Theta_{\text{alg}}^{< R} \nonumber\\
& \quad \times Z_g^{(0)}(s, x\omega_J) Z_g^{(0)} {\left(s, (1-x)\omega_J\right)}\,.
\end{align}

The constraint in Eq.~\eqref{eq:antiIn} leads to the following $k_{\perp}$ integral~\cite{Kang:2016mcy}
\begin{align}\label{eq:antiInteIn}
  \int \frac{\mathrm{d} k_{\perp}}{k_{\perp}^{1+2\epsilon}} \Theta_{\text{alg}}^{< R} 
    &= \int_{0}^{x(1-x)\omega_J \tan\frac{\mathcal{R}}{2}} \frac{\mathrm{d} k_{\perp}}{k_{\perp}^{1+2\epsilon}} = -\frac{1}{2\epsilon} {\left( \omega_J \tan \frac{\mathcal{R}}{2} \right)}^{-2\epsilon} {\left[ x(1-x) \right]}^{-2\epsilon} ,
\end{align}
Thus, the NLO contribution is obtained by performing the $\epsilon$-expansion and, after suitable rearrangements, can be written as
\begin{align}\label{eq:NLOpureg}
&\tilde{M}_{g,\text{bare}}^{gg,(1)}(s, \omega_J) \nonumber\\
&=\frac{\alpha_s}{2\pi} {\left[ \frac{C_A}{{\epsilon}^2} + \frac{C_A}{\epsilon}\mathcal{L} + \frac{C_A}{2} \mathcal{L}^2 
    - C_A\frac{\pi^2}{12} \right]} Z_g^{(0)}(s, \omega_J) \nonumber \\
& + \frac{\alpha_s}{2\pi} {\left( -\frac{1}{\epsilon} - \mathcal{L} \right)}  \int_0^1 \mathrm{d} x \frac{ x(1-x)\hat{p}_{gg}(x) Z^{(0)}_g(s,x\omega_J) Z_g^{(0)}(s,(1-x)\omega_J)}{{(1-x)}_+} \nonumber \\
& + \frac{\alpha_s}{2\pi} \int_0^1 \mathrm{d} x  \left[\frac{\ln x}{1-x} + {\left[ \frac{\ln(1-x)}{1-x} \right]}_+\right] 2 x(1-x)\hat{p}_{gg}(x) Z_g^{(0)}(s,x\omega_J) Z_g^{(0)}(s,(1-x)\omega_J).
\end{align}
For the pure-gluon system, we set $n_f=0$. The logarithm is
\begin{align}\label{eq:log}
  \mathcal{L} = \ln \frac{\mu^2}{\omega_J^2 \tan^2 \frac{\mathcal{R}}{2}}\,.
\end{align}
The first line is not yet the structure for the NLO exclusive jet function. There is an inherent ambiguity in the simple factorization $\tilde{M}=J\times Z$. Therefore, we use the constraint that $Z$ must preserve $Z(s=0)=1$ (probability conservation) under radiative correction. To ensure this, we can subtract certain terms from the second and third lines and add them back to the first line.
After this reorganization, we have
\begin{align}\label{eq:gmultiNLO}
  & \tilde{M}_{g,\text{bare}}^{gg,(1)}(s, \omega_J)  \nonumber\\
  &= \frac{\alpha_s}{2\pi} {\left[ \frac{C_A}{{\epsilon}^2} + \frac{C_A}{\epsilon}\mathcal{L} + \frac{\beta_0^{n_f=0}}{2}\frac{1}{\epsilon} + \frac{C_A}{2} \mathcal{L}^2  +\frac{\beta_0^{n_f=0}}{2}\mathcal{L} + d^{g,n_f=0}_{\textrm{anti-$k_T$}} \right]} Z_g^{(0)}(s, \omega_J) \nonumber \\
  & + \frac{\alpha_s}{2\pi} {\left( -\frac{1}{\epsilon} - \mathcal{L} \right)}  \int_0^1 \mathrm{d} x \frac{1}{2} \hat{P}_{gg}(x) Z^{(0)}_g(s,x\omega_J) Z_g^{(0)}(s,(1-x)\omega_J) 
  + Z^{(1), \text{finite}}_g\,.
\end{align}
In this way, all the corrections to $Z$ will vanish for $Z(s=0)$. Thus, if we start with a properly normalized $Z(s=0)=1$ distribution, the normalization is always guaranteed after radiative correction. Such reorganizations also give the correct coefficients $d^{g,n_f=0}_{\textrm{anti-$k_T$}} = C_A\left(\frac{67}{9}-\frac{3\pi^2}{4}\right)$ for anti-$k_T$ exclusive jet~\cite{Kang:2016mcy}. The first term contains single and double poles as corrections to the exclusive jet function. In the second term, the single pole comes from a collinear divergence of infrared origin. The last term is an NLO leftover without any logarithmic enhancement, which is denoted as 
\begin{subequations}\label{eq:Z1mupureg}
\begin{align}
  Z^{(1),\text{finite}}_g(s,\omega_J) &= \frac{\alpha_s}{2\pi} \int_0^1 \mathrm{d} x  {\left[f_{g}^{gg}(x)\right]}_+ Z_g^{(0)}(s,x\omega_J) Z_g^{(0)}(s,(1-x)\omega_J),\\
f_{g}^{gg}(x) &= \left[\frac{\ln x}{1-x} + {\left( \frac{\ln(1-x)}{1-x} \right)}_+\right] 2 x(1-x)\hat{p}_{gg}(x).
\end{align}
\end{subequations}

Finally, we obtain the full bare exclusive jet multiplicity generating function at NLO and reorganize them into the factorized structure of exclusive jet function times the multiplicity generating function
\begin{align}\label{eq:bareJetMulti}
  \tilde{M}^g_{g, \text{bare}}(s, \omega_J) 
  &= \tilde{M}_g^{g,(0)}(s, \omega_J) + \tilde{M}^{gg,(1)}_{g, \text{bare}}(s, \omega_J) \nonumber \\
  &= \left[J_g^{(0)}+J_{g}^{(1)} \right]\left[Z^{(0)}_g(s, \omega_J) + Z^{(1)}_g(s,\omega_J)\right]+\mathcal{O}(\alpha_s^2), 
\end{align}
where the NLO exclusive jet function and the NLO multiplicity generating function are given by
\begin{align}\label{eq:bareJ}
  J_g^{(0)}+J_{g}^{(1)} &= 1+ \frac{\alpha_s}{2\pi} {\left[ \frac{C_A}{{\epsilon}^2} + \frac{C_A}{\epsilon}\mathcal{L} + \frac{\beta_0^{n_f=0}}{2}\frac{1}{\epsilon} + \frac{C_A}{2} \mathcal{L}^2  +\frac{\beta_0^{n_f=0}}{2}\mathcal{L} + d^{g,n_f=0}_{\textrm{anti-$k_T$}} \right]}\,, \\
  \label{eq:bareZ}
  Z^{(0)}_g + Z^{(1)}_g &= Z_g^{(0)}(s, \omega_J)+ Z^{(1), \text{finite}}_g(s,\omega_J)\nonumber\\
  & + \frac{\alpha_s}{2\pi} {\left( -\frac{1}{\epsilon} - \mathcal{L} \right)}  \int_0^1 \mathrm{d} x \frac{1}{2} \hat{P}_{gg}(x) Z^{(0)}_g(s,x\omega_J) Z_g^{(0)}(s,(1-x)\omega_J) \,.
\end{align}
In the derivation from Eq.~(\ref{eq:NLOpureg}) to Eq.~(\ref{eq:antiInteIn}), all the poles for $J_g$ and $Z_g$ are of IR nature. However, once we include the scaleless integral, corresponding to the rightmost diagram in Fig.~\ref{fig:g2gg-exclusive}, the poles in $J_g$ will be altered to UV nature, while those in $Z_g$ still remain to be IR poles.

\paragraph{Soft Sensitivity of $Z_g$}
We note that our choice of a physical boundary condition in which $Z \rightarrow 1$ when $x$ or $(1-x)$ becomes comparable to or smaller than $\Lambda_{\rm QCD}/\omega$, ensures that the $x$-integration in Eq.~\eqref{eq:bareZ} remains finite. However, due to the inherent soft sensitivity of the multiplicity observable, large soft logarithms are still expected near the endpoints of the $x$-integration. Indeed, if one expands $Z$ in powers of the small parameter $\Lambda_{\rm QCD}/\omega$, these logarithmic divergences become explicit. The resummation of these soft logarithms will be addressed using angular ordering in Sec.~\ref{sec:angularOrdering}. We have chosen this more physical boundary condition to facilitate numerical checks of different nonperturbative models for $Z_g$ and to systematically investigate their phenomenological impact.

\subsection{Renormalization}\label{sec:renormalization}
The bare jet multiplicity generating function in Eq.~\eqref{eq:bareJetMulti} contains both UV divergences and IR divergences. From the definition in Eq.~\eqref{eq:exclusiveJetMultiDefinition} and also its factorized form, $\tilde{M}_{g,\text{bare}}^{g} = J_{g}^{\text{bare}} \times Z_{g}^{\text{bare}}$, a renormalization strategy is suggested, where the UV divergences of each factor are removed independently to obtain the renormalized quantity, $\tilde{M}_{g}^g = J_g \times Z_g$. This procedure then leads to renormalization group equations (RGEs) for the renormalized $J_g$ and $Z_g$. The explicit forms of the bare functions are already given in Eqs.~(\ref{eq:bareJ}) and (\ref{eq:bareZ}). 

\paragraph{Renormalization and evolution of $J_g$} The exclusive jet function contains the UV pole and is renormalized in a multiplicative manner
\begin{align}
J_g &= \mathcal{Z}_{J_g}^{-1}J_g^{\rm bare}  =\mathcal{Z}_{J_g}^{-1}\left(J_g^{(0)}+J_g^{(1)}\right), \\
\mathcal{Z}_{J_g} & = 1 + \frac{\alpha_s(\mu^2)}{2\pi} {\left[ \frac{C_A}{{\epsilon}^2} + \frac{C_A}{\epsilon}\mathcal{L} + \frac{\beta_0^{n_f=0}}{2}\frac{1}{\epsilon} \right]}.
\end{align}

Employing the LO relation $\frac{\rm d\alpha_s(\mu^2)}{\rm d\ln\mu^2} = -\epsilon\alpha_s(\mu^2)$, the corresponding anomalous dimension is derived from the
\begin{align}
\gamma_{J_g} &= \frac{\mathrm{d}\ln J_g}{\rm d\ln\mu^2} = -\frac{\mathrm{d} \ln \mathcal{Z}_{J_g}}{\rm d\ln\mu^2} \\
&= \frac{\rm d}{\rm d\ln\mu^2}\frac{\alpha_s(\mu^2)}{2\pi}\left[ \frac{C_A}{{\epsilon}^2} + \frac{C_A}{\epsilon}\mathcal{L} + \frac{\beta_0^{n_f=0}}{2}\frac{1}{\epsilon} \right]\\
&= -\frac{\alpha_s(\mu^2)}{2\pi} \left(C_A\mathcal{L}+\frac{\beta_0^{n_f=0}}{2}\right).
\end{align}
The initial scale is chosen to be the jet scale $\mu_J = \omega_J \tan \frac{\mathcal{R}}{2}$ for which the fixed-order logarithms vanish, $\mathcal{L}=0$. 
At this scale, the jet function satisfies
\begin{align}
J_g(\omega_J, R, \mu_J) 
= 1 + \frac{\alpha_s(\mu_J^2)}{2\pi}\,
   d^{g,n_f}_{\textrm{anti-$k_T$}},
\end{align}
where $d^{g,n_f}_{\textrm{anti-$k_T$}}$ denotes the finite NLO constant arising from the anti-$k_T$ jet algorithm.

The scale dependence of the jet function is governed by the anomalous dimension
\begin{align}
\frac{\mathrm{d}}{\mathrm{d}\ln\mu^2} J_g(\omega_J,\mu) 
= -\frac{\alpha_s(\mu^2)}{2\pi} 
  \left( C_A \mathcal{L} + \frac{\beta_0^{n_f=0}}{2} \right),
\end{align}
which evolves $J_g(\omega_J,\mu)$ from the jet scale $\mu_J$ 
to the hard scale $\mu \sim \omega_J$.

\paragraph{Renormalization and evolution of $Z_g$} In the $\overline{\rm MS}$ renormalization prescription, the counterterm only subtracts the pole terms in $\epsilon$.
Because of the non-linear structure in the NLO correction to $Z_g$, this is actually a non-linear functional ($Z$-dependent) renormalization procedure. The bare quantity is related to the renormalized $Z_g$ by
\begin{align}
Z^{\rm bare} &= Z_g(s,\omega_J,\mu) - \frac{\alpha_s(\mu^2)}{2\pi} \frac{1}{\epsilon}\int_0^1 \mathrm{d} x \frac{1}{2} \hat{P}_{gg}(x) Z^{(0)}_g(s,x\omega_J) Z_g^{(0)}(s,(1-x)\omega_J) \\
 &\approx Z_g(s,\omega_J, \mu) - \frac{\alpha_s(\mu^2)}{2\pi} \frac{1}{\epsilon}\int_0^1 \mathrm{d} x \frac{1}{2} \hat{P}_{gg}(x) Z_g(s,x\omega_J,\mu) Z_g(s,(1-x)\omega_J,\mu)\, ,
\end{align} 
where in the second line, we have replaced $Z_g^{(0)}$ by the renormalized $Z_g$ on the right hand side of the equation, up to differences of order $\alpha_s^2$. 
Finally, we take the derivative with respect to $\ln \mu^2$ on both sides and use the scale independence of $Z^{\rm bare}$, and the LO relation $\frac{\partial \alpha_s(\mu^2)}{\partial \ln \mu^2} = -\epsilon\alpha_s(\mu^2)$ to get the RG equation for $Z_g$
\begin{align}
\frac{\partial}{\partial \ln\mu^2} Z_g(s,\omega_J,\mu) &=  -\frac{\alpha_s(\mu^2)}{2\pi} \int_0^1 \mathrm{d} x \frac{1}{2} \hat{P}_{gg}(x) Z_g(s,x\omega_J,\mu) Z_g(s,(1-x)\omega_J,\mu)\, ,
\end{align}
with an initial condition at $\mu=\omega_J \tan \frac{\mathcal{R}}{2}$ being $Z_g^{(0)}+Z_g^{(1),\rm finite}$.
In practice, however, since we always parameterize $Z_g$ at the boundary near $\mu\sim Q_0$, we will use the \textit{backward evolution scheme} of the equation and evolve $Z_g$ from $Q_0$ to $\omega_J \tan \frac{\mathcal{R}}{2}$,
\begin{align}
\frac{\partial}{\partial \ln\mu^2} Z_g(s,\omega_J,\mu) &=  \frac{\alpha_s(\mu^2)}{2\pi} \int_0^1 \mathrm{d} x \frac{1}{2} \hat{P}_{gg}(x) Z_g(s,x\omega_J,\mu) Z_g(s,(1-x)\omega_J,\mu)\,
\end{align}
where the sign of the evolution is reversed, and $\mu$ starts from $\mu\sim Q_0$ and evolve up to scale $\mu\sim \omega_J \tan \frac{\mathcal{R}}{2}$. In this scheme, the finite term $Z_g^{(1),\rm finite}$ is evaluated using the evolved $Z_g$, and is added to the final result after the evolution to the jet scale $\mu_J =\omega_J\tan \frac{\mathcal{R}}{2}$.

\subsection{Including the quarks}\label{sec:exclusiveMulti} 
The inclusion of the quark case follows exactly the same procedure as in the pure-gluon case. We will just list the final bare results here and perform the renormalization procedure accordingly.
For brevity, we use the following abbreviation
\begin{align}
  Z_j^{(0)}(x) \cdot Z_k^{(0)}(1-x) \equiv Z_j^{(0)}(s, x\omega_J) Z_k^{(0)} {\left(s, (1-x)\omega_J \right)}, \label{eq:saveSpace}
\end{align}
Then, for $q\rightarrow q+g$
\begin{align}
\tilde{M}^{qg,(1)}_{q, \text{bare}}(s,  \omega_J)
&= \frac{\alpha_s}{2\pi}  {\left[ \frac{C_F}{{\epsilon}^2} + \frac{C_F}{\epsilon}\mathcal{L} + \frac{C_F}{2}\mathcal{L}^2 +\frac{3C_F}{2 \epsilon} + \frac{3C_F}{2}\mathcal{L} + d^{q}_{\textrm{anti-$k_T$}} \right]} Z_q^{(0)}(s, \omega_J)  \nonumber \\
& + \frac{\alpha_s}{2\pi} {\left( -\frac{1}{\epsilon} - \mathcal{L} \right)} \int \mathrm{d} x \hat{P}_{qq}(x) Z_q^{(0)}(x) \cdot Z_g^{(0)}(1-x) \nonumber \\
&+ \frac{\alpha_s}{2\pi} \int \mathrm{d} x  {\left[ f_{q}^{qg}(x) \right]}_+ Z^{(0)}_q(x) \cdot Z^{(0)}_g(1-x) 
\end{align}
with
\begin{align}
  f_{q}^{qg}(x) 
  &= 2 C_F \frac{1+x^2}{1-x} \ln x  + 2C_F( 1 + x^2 ) {\left[ \frac{\ln (1-x)}{1-x} \right]}_+ +  C_F (1-x) \\
d^{q}_{\textrm{anti-$k_T$}} &= C_F\left(\frac{13}{2}-\frac{3\pi^2}{4}\right)
\end{align}
For the sum of $g\rightarrow gg$ and $g\rightarrow q\bar{q}$
\begin{align}
    \tilde{M}_{g, \text{bare}}^{gg+q\bar{q},(1)}(s, \omega_J)
    & = \frac{\alpha_s}{2\pi} {\left[ \frac{C_A}{{\epsilon}^2} + \frac{C_A}{\epsilon}\mathcal{L} + \frac{C_A}{2} \mathcal{L}^2 + \frac{\beta_0^{n_f}}{2 \epsilon} + \frac{\beta_0^{n_f}}{2} \mathcal{L} + d^{g,n_f}_{\textrm{anti-$k_T$}}\right]} Z_g^{(0)}(s, \omega_J) \nonumber \\
    & + \frac{\alpha_s}{2\pi} {\left( -\frac{1}{\epsilon} - \mathcal{L} \right)}  \int \mathrm{d} x \bigg\{ \frac{1}{2} \hat{P}_{gg}(x) Z^{(0)}_g(x)\cdot Z_g^{(0)}(1-x) \nonumber\\
    & \hskip4cm+ n_f \hat{P}_{qg}(x)  Z^{(0)}_q(x)\cdot Z_q^{(0)}(1-x) \bigg\}\nonumber \\
    &+ \frac{\alpha_s}{2\pi} \int \mathrm{d} x {\left[f^{gg}_g(x)\right]}_+Z_g^{(0)}(x) \cdot Z_g^{(0)}(1-x) \nonumber\\
    &+ \frac{\alpha_s}{2\pi} \int \mathrm{d} x f_{g}^{q\bar{q}}(x) \left[ Z^{(0)}_q(x)\cdot Z_q^{(0)}(1-x) - Z_g^{(0)}(s,\omega_J) \right]
\end{align}
with 
\begin{align}
  f_{g}^{q\bar{q}}(x) &= 2n_f \left[\hat{P}_{qg}(x)\ln[x(1-x)] + T_F x(1-x)\right],\\
d^{g,n_f}_{\textrm{anti-$k_T$}} &=  C_A\left(\frac{67}{9}-\frac{3\pi^2}{4}\right) - T_F n_f\frac{23}{9}
\end{align}
Since both daughter partons are inside the jet and each contributes to the multiplicity, we must account for the full final-state configuration. For the quark jets, the channels for $q \to q+g$ and $q \to g+q$ correspond to the same physical configuration. To write them explicitly without double counting, we can use the symmetry of the splitting function $P_{qq}(x) = P_{gq}(1-x)$, and include an additional factor $1/2$. We treat the quark and anti-quark multiplicity generating function as equal, i.e.\ $Z_{\bar{q}} \equiv Z_q$. 

From these results, we can extract the LO and NLO contribution along with the LL evolution equations as the last section shows. Here, we only show the evolution equations for the generating functions, presented in the \textit{backward evolution scheme},
\begin{subequations}\label{eq:muEvolution}
\begin{align}
\frac{\partial Z_g(s,\omega_J, \mu)}{\partial \ln \mu^2}
&=\frac{\alpha_s(\mu^2 )}{2\pi}\int_0^1 \mathrm{d} x   
\bigg\{ \frac{1}{2} \hat{P}_{gg}(x) Z_g(s,x\omega_J, \mu) Z_g(s,(1-x)\omega_J, \mu)  \nonumber\\
&\quad + n_f \hat{P}_{qg}(x) Z_{q}(s, x\omega_J, \mu)Z_{q}(s, (1-x)\omega_J, \mu) \bigg\}, \\
\frac{\partial Z_q(s,\omega_J,\mu)}{\partial \ln\mu^2}
&= \frac{\alpha_s(\mu^2 )}{2\pi} \int_0^1 \mathrm{d} x \hat{P}_{qq}(x) Z_q(s,x\omega_J,\mu) Z_g(s,(1-x)\omega_J,\mu)
\end{align}
\end{subequations}
After evolution, the NLO correction added to the evolved generating function is 
\begin{align}
Z_{i}(s, \omega_J, \mu_J) = Z_{i}^{(0)}(s, \omega_J, \mu_J) + \frac{\alpha_s(\mu^2_J)}{2\pi} Z_{i}^{(1)}(s, \omega_J, \mu_J)
\end{align}
with
\begin{subequations}\label{eq:NLOcoefficientmu}
\begin{align}
Z_{g}^{(1)}(s, \omega_J, \mu) 
&= \int_0^1 \mathrm{d} x f^{q\bar{q}}_g(x) \left[ Z_{q}^{(0)}(s, x\omega_J, \mu)Z_{q}^{(0)}(s, (1-x)\omega_J, \mu) - Z^{(0)}_g(s,\omega_J,\mu) \right] \nonumber\\
&+  \int_0^1 \mathrm{d} x {\left[f^{gg}_g(x)\right]}_+ Z_{g}^{(0)}(s, x\omega_J, \mu)Z_{g}^{(0)}(s, (1-x)\omega_J, \mu) \\
Z_{q}^{(1)}(s, \omega_J, \mu) 
&= \int_0^1 \mathrm{d} x {\left[f^{qg}_q(x)\right]}_+ Z_{q}^{(0)}(s, x\omega_J, \mu)Z_{g}^{(0)}(s, (1-x)\omega_J, \mu) 
\end{align}
\end{subequations}

\subsection{Implementing the angular ordering}\label{sec:angularOrdering}

If one naively applies the original set of evolution equations, the resulting multiplicity evolves extremely rapidly toward large fluctuations. The reason is that the multiplicity distribution is highly sensitive to soft radiation, so it is important to treat the endpoints \(x\to 0,1\) carefully in the non-linear convolution of the generating functional \(Z\). For this purpose,  we follow the approach of Bassetto \emph{et al.}~\cite{Bassetto:1983mvz}, who showed that implementing angular ordering is sufficient to improve the evolution at LL level when using the soft approximation of the gluon splitting function. This angular evolution scheme encodes a key physical insight of QCD:~because of destructive interference, soft gluon emissions are ordered in angle rather than in energy. Incorporating this coherence improves multiplicity predictions in the soft sector and underlies modern parton-shower algorithms and resummation formalisms~\cite{MARCHESINI19841,Catani:1990rr,Banfi:2006gy,Bahr:2008pv,Bewick:2019rbu}.

To implement angular ordering, one chooses the quantity $\zeta\equiv 1-\cos\theta$ as the evolution variable, where $\theta$ denotes the emission angle. The transverse momentum is then $k_\perp^2 = 2\,\zeta\,{[x(1-x)\,\omega_J]}^2$ and the running coupling is evaluated at this scale, $\alpha_s(k_{\perp}^2)$. The evolution is restricted to emissions with $k_\perp^2>Q_0^2$, avoiding the nonperturbative regime~\cite{Bassetto:1983mvz}. The angular-ordered evolution therefore resums the logarithms associated with large-angle hierarchies and ensures soft emissions are included in the correct kinematic regime. The improved evolution equation (\textit{backward evolution scheme}) reads
\begin{subequations}\label{eq:zetaEvolution}
\begin{align}
\frac{\partial Z_g(s,\omega_J, \zeta)}{\partial \ln \zeta}
&=\int_0^1 \mathrm{d} x \frac{\alpha_s(k_{\perp}^2 )}{2\pi} \Theta(k_{\perp}^2 > Q_0^2)  
\bigg\{ \frac{1}{2} \hat{P}_{gg}(x) Z_g(s,x\omega_J, \zeta) Z_g(s,(1-x)\omega_J, \zeta)  \nonumber\\
&\quad + n_f \hat{P}_{qg}(x) Z_{q}(s, x\omega_J, \zeta)Z_{q}(s, (1-x)\omega_J, \zeta) \bigg\}, \\
\frac{\partial Z_q(s,\omega_J,\zeta)}{\partial \ln\zeta}
&=  \int_0^1 \mathrm{d} x \frac{\alpha_s(k_{\perp}^2 )}{2\pi} \Theta(k_{\perp}^2 > Q_0^2) \hat{P}_{qq}(x) Z_q(s,x\omega_J,\zeta) Z_g(s,(1-x)\omega_J,\zeta)\,.
\end{align}
\end{subequations}

Furthermore, with $\mu^2$ chosen to be $\zeta\omega_J^2$, there is also a subtle change to the finite NLO leftover $Z^{(1)}$ terms appearing in Eq.~\eqref{eq:NLOcoefficientmu}. Taking $Z_g$ in the pure-gluon system as shown in Eq.~\eqref{eq:NLOpureg} as an example, tracing its origin, $Z_g^{(1)}$ comes from the $\epsilon$-expansion of
\begin{align}
&-\mu^{2\epsilon}\frac{1}{2\epsilon} {\big(\omega_J \tan\tfrac{\mathcal{R}}{2}\big)}^{-2\epsilon}{[x(1-x)]}^{-2\epsilon} \hat{P}_{gg}(x, \epsilon) \nonumber\\
&= -\mu^{2\epsilon}\frac{1}{2\epsilon} {\big(\omega_J \tan\tfrac{\mathcal R}{2}\big)}^{-2\epsilon} C_A 2{(1-x(1-x))}^2
\left[ \frac{1}{{(1-x)}^{1+2\epsilon}x^{2\epsilon}} + \frac{1}{x^{1+2\epsilon}{(1-x)}^{2\epsilon}} \right].
\end{align}
Expanding in $\epsilon$ and keeping the pieces that contribute to $Z^{(1)}$ as $\epsilon\to 0$ yields terms proportional to
\begin{equation}
  2C_A 2{(1-x(1-x))}^2 \left( {\left[\frac{\ln(1-x)}{1-x}\right]}_+ + \frac{\ln x}{1-x} + \{x\leftrightarrow 1-x\} \right),
\end{equation}
which gives the same structures appearing in the function $f^g_{gg}$ defined in Eq.~\eqref{eq:Z1mupureg}. The key observation is that $\mu^2$ in the original derivation with dimensional regularization effectively acts as an infrared regulator for transverse momentum integrals $k_{\perp}^2$, rather than the regulator of emission angles. If instead we parameterize the regulator in angular variables, e.g. $\mu \simeq \theta_\mu\, x(1-x)\,\omega_J$ and interpret $\theta_\mu$ as an angular separation regulator, then the factor $\mu^{2\epsilon}\propto {[x(1-x)]}^{2\epsilon}$ cancels the ${[x(1-x)]}^{-2\epsilon}$ factor  coming from the transverse momentum integral with the jet algorithm constraint in Eq.~\eqref{eq:antiInteIn}. Consequently, the $\mathcal O(\epsilon)$ piece that generated $Z_g^{(1)}$ disappears in this angular-regulated scheme. This is not a mathematical sleight-of-hand but follows from using a physically motivated infrared cutoff aligned with the coherent, angular-ordered radiation pattern of QCD. With quarks taken into account, it is not a complete cancellation, and the finite NLO leftovers that are compatible with the angular ordering are
\begin{subequations}\label{eq:NLOcoefficientzeta}
\begin{align}
Z_{g}^{(1),{\rm finite}}(s, \omega_J, \zeta) 
&= n_f\times \int_0^1 dx \frac{\alpha_s(k_{\perp}^2)}{2\pi} \Theta(k_{\perp}^2 > Q_0^2) T_F[2x(1-x)] \nonumber
\\ & \quad  \times   \left[ Z_{q}^{(0)}(s, x\omega_J, \zeta) Z_{q}^{(0)}(s, (1-x)\omega_J, \zeta) - Z^{(0)}_g(s,\omega_J,\zeta) \right]\, \\
Z_{q}^{(1),{\rm finite}}(s, \omega_J, \zeta) 
&= \int_0^1 dx  \frac{\alpha_s(k_{\perp}^2)}{2\pi} \Theta(k_{\perp}^2 > Q_0^2) C_F(1-x)  \nonumber 
\\ & \quad \times \left[ Z_{q}^{(0)}(s, x\omega_J, \zeta) Z_{g}^{(0)}(s, (1-x)\omega_J, \zeta) - Z_q^{(0)}(s, \omega_J, \zeta) \right]\,.
\end{align}
\end{subequations}
Apparently, they still satisfy the probability conservation that $Z^{(1)}(s=0) = 0$.

To consistently combine the NLO finite corrections with the backward angular ordered evolution, we first specify a nonperturbative initial condition for the generating functional $Z$ at a small angle $\theta_0$ as discussed in Sec.~\ref{sec:NPModel}. Next, we evolve $Z$ from $\theta_0$ up to the jet radius $R$ using the coupled equations in Eqs.~\eqref{eq:zetaEvolution}. Once the resummed solution $Z^{(0)}$ is obtained, we evaluate the NLO contributions according to Eqs.~\eqref{eq:NLOcoefficientzeta}. The full generating functional is $Z_i \;=\; Z_i^{(0)} \,+\, Z_i^{(1),~\rm finite}$. This procedure ensures that both the angular-ordered resummation and the finite NLO corrections are consistently incorporated.

\subsection{Nonperturbative modeling of the generating function}\label{sec:NPModel}

\begin{figure}
    \centering
    \includegraphics[height=0.55\textwidth]{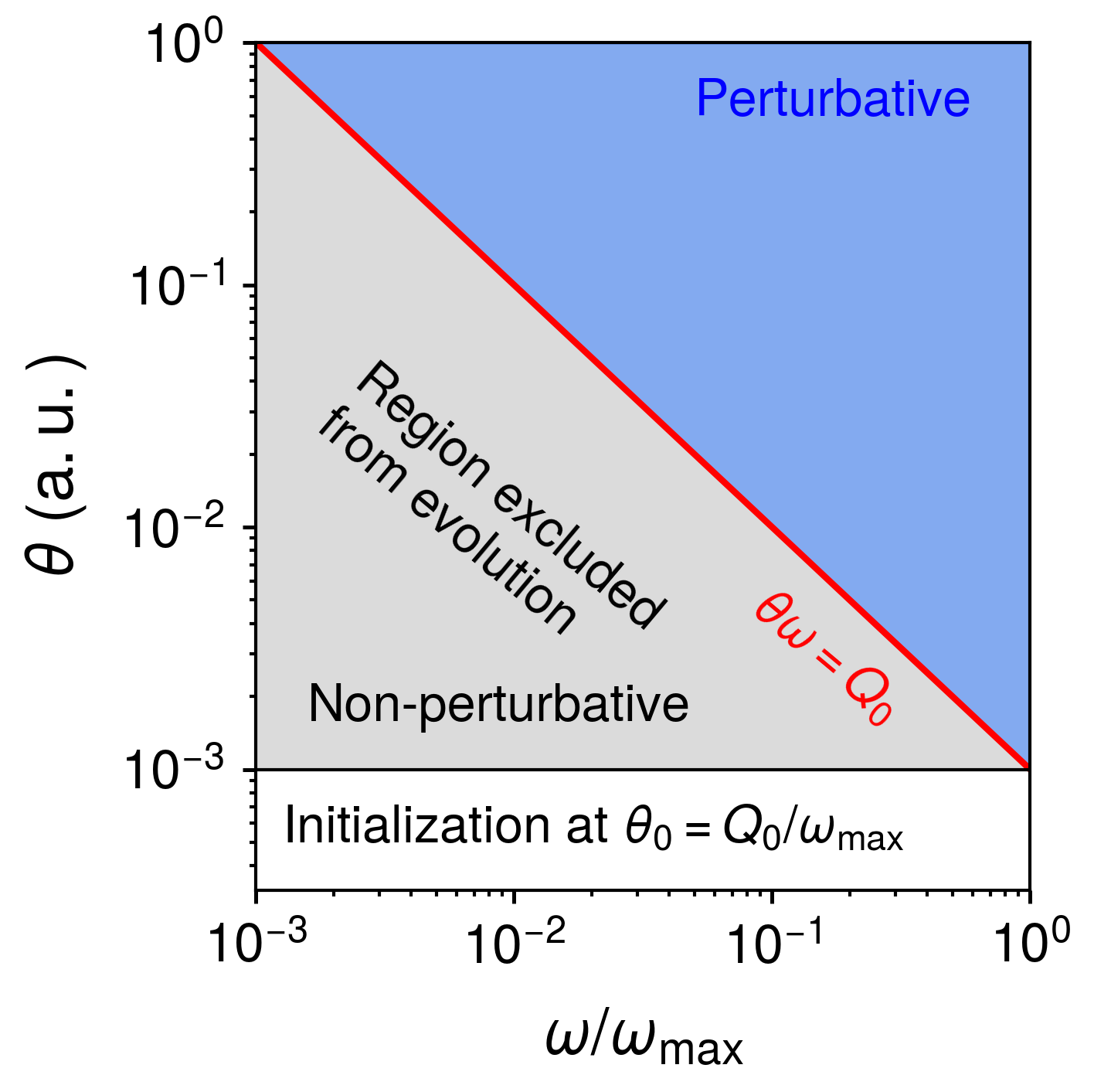}
    \caption{On the log-log plot of energy and angle, evolution equations are initialized at an angle that determined by the largest jet energy under consideration. Kinematics within the gray shaded region is removed from the evolution such that the initial condition is preserved at the boundary $\theta\omega = Q_0$.}\label{fig:IC}
\end{figure}

The multiplicity distribution function $P_i(n,\omega,\mu)$ is not infrared- or collinear-safe and therefore contains an intrinsic nonperturbative (NP) component that cannot be computed within perturbative QCD, as we discussed in Sec.~\ref{sec:renormalization}. To proceed, we model its functional form at an initial scale. Since the evolution equations in Eq.~\eqref{eq:zetaEvolution} are formulated in the angular ordering framework~\cite{Bassetto:1983mvz}, we specify the initial condition on a reference curve in the $(\omega,\theta)$ plane. The key insight is to recognize that the natural scale for the evolution is the transverse momentum $k_\perp \sim \theta \omega$. To ensure that the initial condition provides a genuinely NP input, we choose it along the curve
\begin{align}
  \theta\,\omega = Q_0 ,
\end{align}
where $Q_0 \sim \Lambda_{\rm QCD}$ is a fixed NP cutoff. This choice is illustrated by the red line in Fig.~\ref{fig:IC}. For a chosen maximum jet energy $\omega_{\max}$, this fixes the reference angle as
\begin{align}
  \theta_0 = \frac{Q_0}{\omega_{\max}}.
\end{align}
Thus, for jets at the maximal energy the evolution starts exactly at the NP–perturbative boundary $k_\perp \sim Q_0$. For lower jet energies $\omega < \omega_{\max}$, the initial scale satisfies $\theta_0 \omega < Q_0$, corresponding to the gray-shaded region below the red line in Fig.~\ref{fig:IC}. This is handled self-consistently by the kinematic constraint $\Theta(k_\perp^2 > Q_0^2)$ in the evolution equation as in Eq.~\eqref{eq:zetaEvolution}. This theta function acts as an infrared cutoff, effectively freezing the evolution until the angular variable $\zeta$ becomes large enough that $k_\perp^2 \sim 2\zeta \omega^2 > Q_0^2$. Thus, for the low-energy region, the multiplicity distribution remains at its nonperturbative initial condition, which is the physically expected behavior.

The parameterization of the initial condition at $\theta_0$ consists of two parts. First, we assume that the multiplicity distribution of parton with virtuality $Q_0$ follows a binomial distribution (or, for simplicity, a Poisson distribution)
\begin{align}
P^{\rm IC}(n,\omega, Q_0) =
\begin{cases}
  \frac{1}{n!} {\left[\langle n \rangle (\omega, Q_0)\right]}^n \, e^{- \langle n\rangle(\omega, Q_0)} & \textrm{Poisson}\,, \\[2ex]
\binom{n_{\max}}{n} {\left(\frac{\langle n\rangle(\omega, Q_0)}{n_{\max}}\right)}^n {\left(1-\frac{\langle n\rangle(\omega, Q_0)}{n_{\max}}\right)}^{n_{\max}-n} & \textrm{Binomial}\,.
\end{cases}
\end{align}
Here, $\langle n \rangle(\omega, Q_0)$ is the average number of charged particles at the initial scale, which may depends on $\omega$ and $Q_0$. This is the only information needed to build the Poisson model. For the binomial model, we need an additional parameter which is the maximum number of particles $n_{\max}\geq \langle n \rangle$ that can be produced from the hadronization of a particle with scale $Q_0$. This is actually more physical, because hadrons have mass, thus there must be a maximum number of hadrons.

Making the Laplace transformation, the initial condition for the multiplicity generating function takes the form
\begin{align}
\label{eq:NP-IC-Z}
    Z^{\rm IC}(s, \omega, Q_0) =
    \begin{cases}
    \exp\left[ \langle n \rangle(\omega, Q_0) \left( e^{-s} - 1 \right) \right] &\textrm{Poisson}\,,\\[2ex]
  {\left[1+\left(e^{-s}-1\right)\frac{\langle n \rangle(\omega, Q_0)}{n_{\max}}\right]}^{n_{\max}} &\textrm{Binomial}\,.
    \end{cases}
\end{align}

Finally, we parameterize the average particle multiplicity using a form motivated by parton-hadron duality~\cite{Azimov:1984np,Dokshitzer:1995ev}:
\begin{align}
\label{eq:NP-IC-Navg}
\langle n\rangle(\omega, Q_0) = \frac{n_0}{1 + c Q_0^2/\omega^2}.
\end{align}
This ensures a correct asymptotic behavior: it approaches $n_0$ at high energy ($\omega \gg Q_0$), while providing a smooth suppression near the production threshold ($\omega \sim Q_0$). The parameter $c$ controls the transition rate between these regimes, implementing the expected threshold effect for low-energy partons.

In our setup, we adopt identical initial conditions for both quark- and gluon-initiated jets for simplicity, although in principle they could differ due to the larger color charge of gluons. Both the binomial and Poisson models are initialized for quark- and gluon-initiated jets at the same scale, $Q_0 = 0.224~\mathrm{GeV}$, with average multiplicity $n_0 = 0.8$ and $c = 1$. For the binomial model, a maximum particle number $n_{\max} = 2$ is imposed, whereas the Poisson distribution does not require such a constraint. We will later explore the sensitivity of the multiplicity distributions to these choices of initial conditions.

\section{Results for exclusive jets}\label{sec:resultsExclusive}
\subsection{Mean and distribution of the internal multiplicity of exclusive jets}
\begin{figure}[ht]
    \centering
    \includegraphics[width=1.0\textwidth]{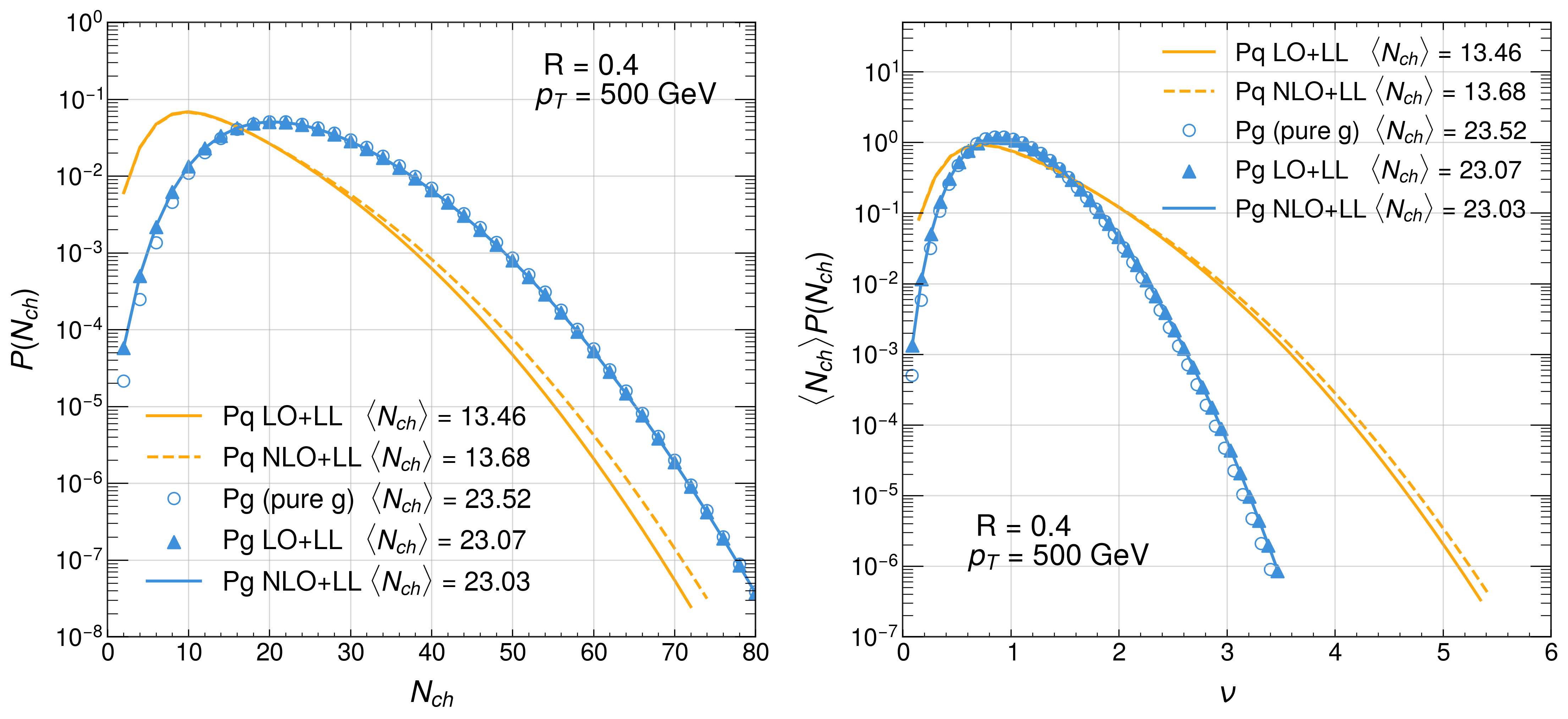}
\caption{Charged-particle multiplicity distributions for exclusive quark and gluon jets with $p_T=500$ GeV and $R=0.4$. 
Left: absolute distributions $P(N_{\rm ch})$; right: rescaled distributions $\langle N_{\rm ch}\rangle P(N_{\rm ch})$ as a function of $\nu = N_{\rm ch}/\langle N_{\rm ch}\rangle$. 
Blue open circles indicate gluon jets in the pure-gluon system. 
In the quark–gluon system, quark jets are shown by orange lines (solid: LO+LL, dashed: NLO+LL) and gluon jets are shown by blue symbols/lines (solid triangles: LO+LL, solid line: NLO+LL).
}\label{fig:exclusivet-jet-Pn}
\end{figure}

Fig.~\ref{fig:exclusivet-jet-Pn} shows the charged-particle multiplicity distributions for exclusive quark and gluon jets with $p_T=500$ GeV and $R=0.4$, comparing LO+LL and NLO+LL results. As expected, gluon jets exhibit a significantly higher mean multiplicity than quark jets, a direct consequence of the larger color factor and the associated enhancement in branching probability. This higher color charge leads to a more intense parton cascade and consequently a denser final state. A notable feature of the distribution is that quark jets dominate the low-multiplicity region, while gluon jets prevail at higher multiplicities, reflecting the underlying color-structure dependence of QCD radiation.

Including NLO corrections to the generating functions, as given in Eqs.~(\ref{eq:NLOcoefficientzeta}), leads to a modest enhancement of quark-jet multiplicity in the high-$N_{\mathrm{ch}}$ tail, whereas the gluon-jet distribution remains nearly unchanged. To isolate the role of $g\to q\bar{q}$ splittings, we also study the “pure-gluon” system (open circles),
where $g\to q\bar{q}$ channels are switched off and all branchings proceed via $g\to gg$.
To make a controlled comparison with the coupled quark-gluon system, we still use the running coupling $\alpha_s$ with three active flavors. 
The resulting mean multiplicity is slightly higher as shown in the legends of Fig.~\ref{fig:exclusivet-jet-Pn}, showing that including quark channels reduces the overall multiplicity because quarks radiate less efficiently than gluons. 
At the same time, including quark channels also increases the probability of low-multiplicity region, because quarks contribute more in this regime.

In the right panel of Fig.~\ref{fig:exclusivet-jet-Pn}, we plot the distribution of the normalized multiplicity $\nu$. The quark and gluon distributions remain clearly distinct, with quark jets exhibiting a broader fluctuation. Moreover, the mixed-system gluon distribution $P_g$ is slightly broader than the pure-gluon $P_g^{\mathrm{pure;g}}$, reflecting additional fluctuations due to quark-gluon mixing. Together, these results demonstrate that multiplicity distributions encode both the flavor of the initiating parton and the detailed branching dynamics.

\begin{figure}[ht]
    \centering
    \includegraphics[width=1.0\textwidth]{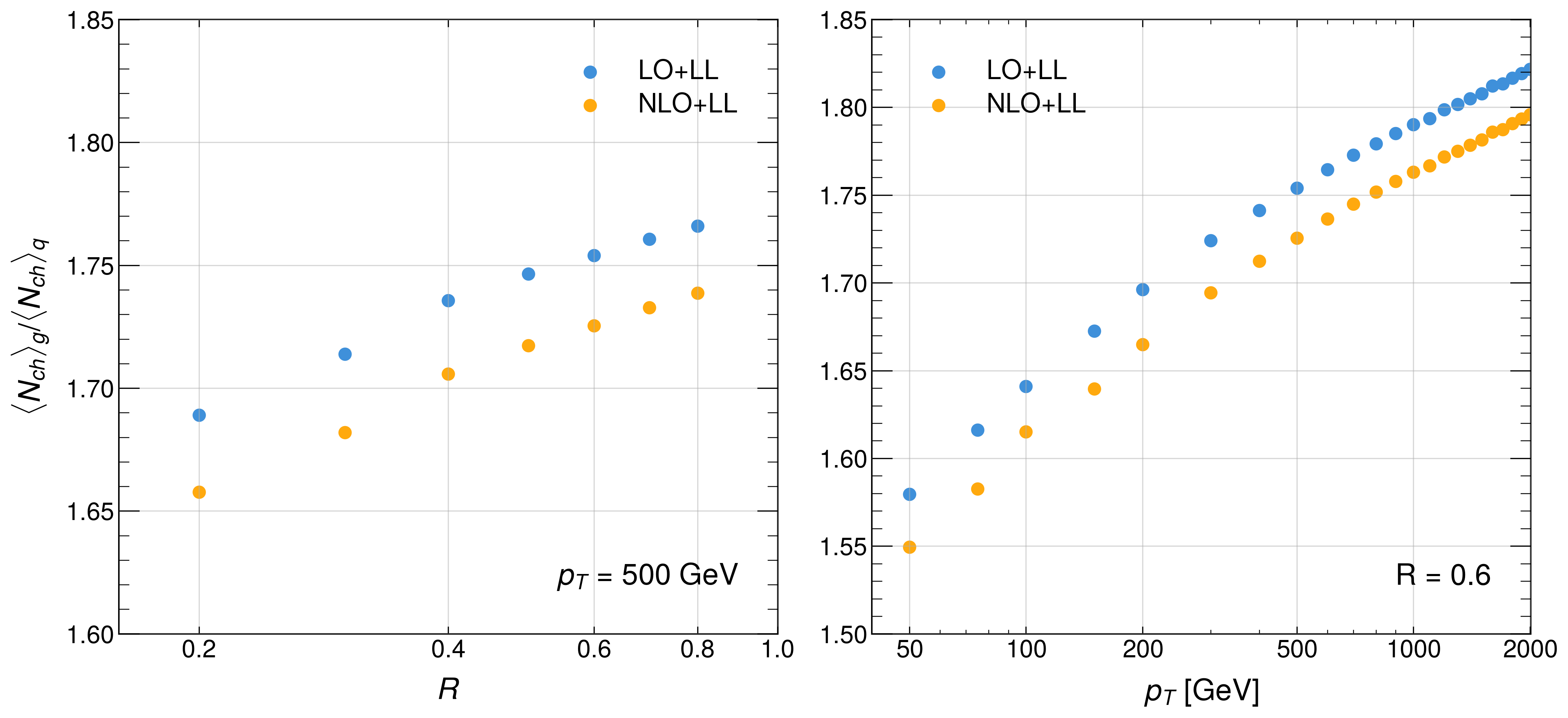}
    \caption{Ratio of the mean charged particle multiplicity between exclusive gluon and quark jets, $\langle N_{ch} \rangle_g / \langle N_{ch} \rangle_q$. Left: dependence on the jet radius $R$ for fixed $p_T = 500$ GeV. Right: dependence on the jet transverse momentum $p_T$ for fixed $R = 0.6$. Results are shown for LO+LL (blue dots) and NLO+LL (orange dots).}
\label{fig:exclusivet_Nq_Ng}
\end{figure}

Fig.~\ref{fig:exclusivet_Nq_Ng} further shows the ratio of the mean charged particle multiplicity in gluon jets to that in quark jets, $\langle N_{\mathrm{ch}} \rangle_g / \langle N_{\mathrm{ch}} \rangle_q$, computed at both LO+LL and NLO+LL accuracy. The ratio exhibits a clear logarithmic dependence on both the jet radius $R$ and the transverse momentum $p_T$, increasing monotonically in both cases. Around $p_T=50$ GeV, the ratio is about 1.55 --- very close to the value needed to explain the LEP data~\cite{ALEXANDER1996659}. The numerical solution of the ratio is not yet approaching the asymptotic value $C_A/C_F$ or its DLA/MLLA improved version~\cite{GAFFNEY1985109,Malaza:1987hj,Dremin:1994bj,Khoze:1996dn}. The NLO+LL prediction is systematically lower than its LO+LL counterpart. This is consistent with our previous finding that NLO corrections predominantly enhance the quark-jet multiplicity, thereby lowering the ratio.

\begin{figure}[ht]
    \centering
    \includegraphics[width=0.55\textwidth]{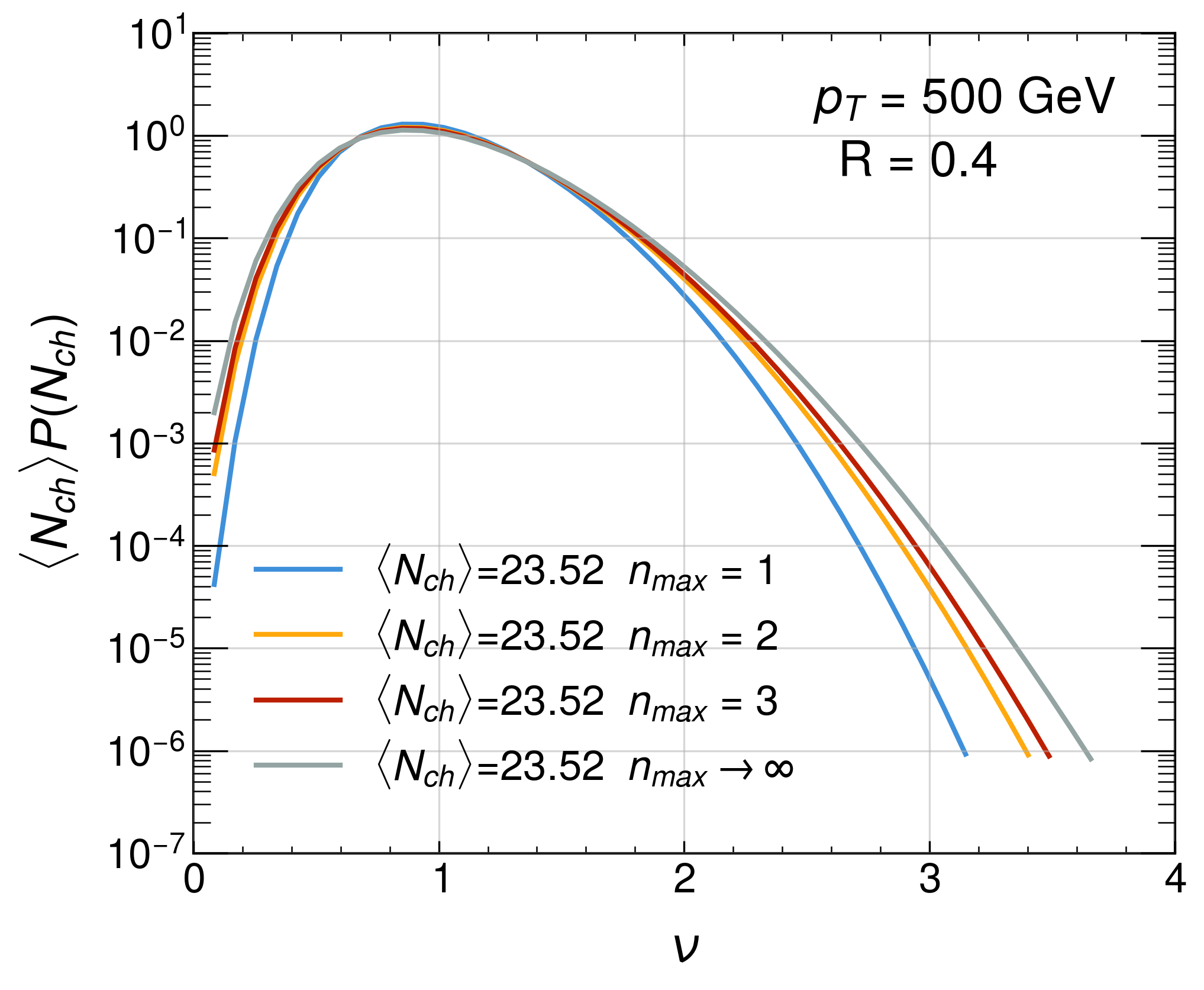}
    \caption{Rescaled charged-particle multiplicity distributions for exclusive gluon jets in the pure-gluon system, plotted as $\langle N_{\rm ch}\rangle P(\nu)$ versus $\nu = N_{\rm ch}/\langle N_{\rm ch}\rangle$ for $p_T=500$ GeV and $R=0.4$. Distributions are shown for different initial binomial parameters $n_{\max} = 1, 2, 3$, with $n_{\max}\to\infty$ corresponding to the Poisson limit.}\label{fig:g2gg-NPmodels}
\end{figure}

In Fig.~\ref{fig:g2gg-NPmodels}, we investigate the sensitivity of the rescaled charged-particle multiplicity distribution to the choice of non-perturbative (NP) initial condition for the gluon generating functional, $Z_g$. We perform this study within a pure-gluon system, where we fix the mean multiplicity $\langle n \rangle(\omega, Q_0)$ at the initial scale and vary the parameter $n_{\max}$ of the binomial NP model; the Poisson limit is recovered as $n_{\max} \to \infty$. A larger $n_{\max}$ corresponds to increased initial fluctuations while the mean multiplicity is held constant. We find that this initial broadening propagates through the perturbative evolution, resulting in a broader final multiplicity distribution. Since the distribution vanishes at $\nu=0$ and for $\nu \gg 1$, the effect of this broadening is most visually pronounced in these extreme regions. We therefore conclude that the behavior of jet multiplicity distributions in the extreme regions of $\nu \ll 1$ and $\nu \gg 1$ is not governed solely by perturbative branching dynamics but also retains a significant sensitivity to the non-perturbative initial conditions.

\subsection{On the KNO scaling and violations in exclusive jet} 

\begin{figure}[ht]
    \centering
    \includegraphics[width=1.0\textwidth]{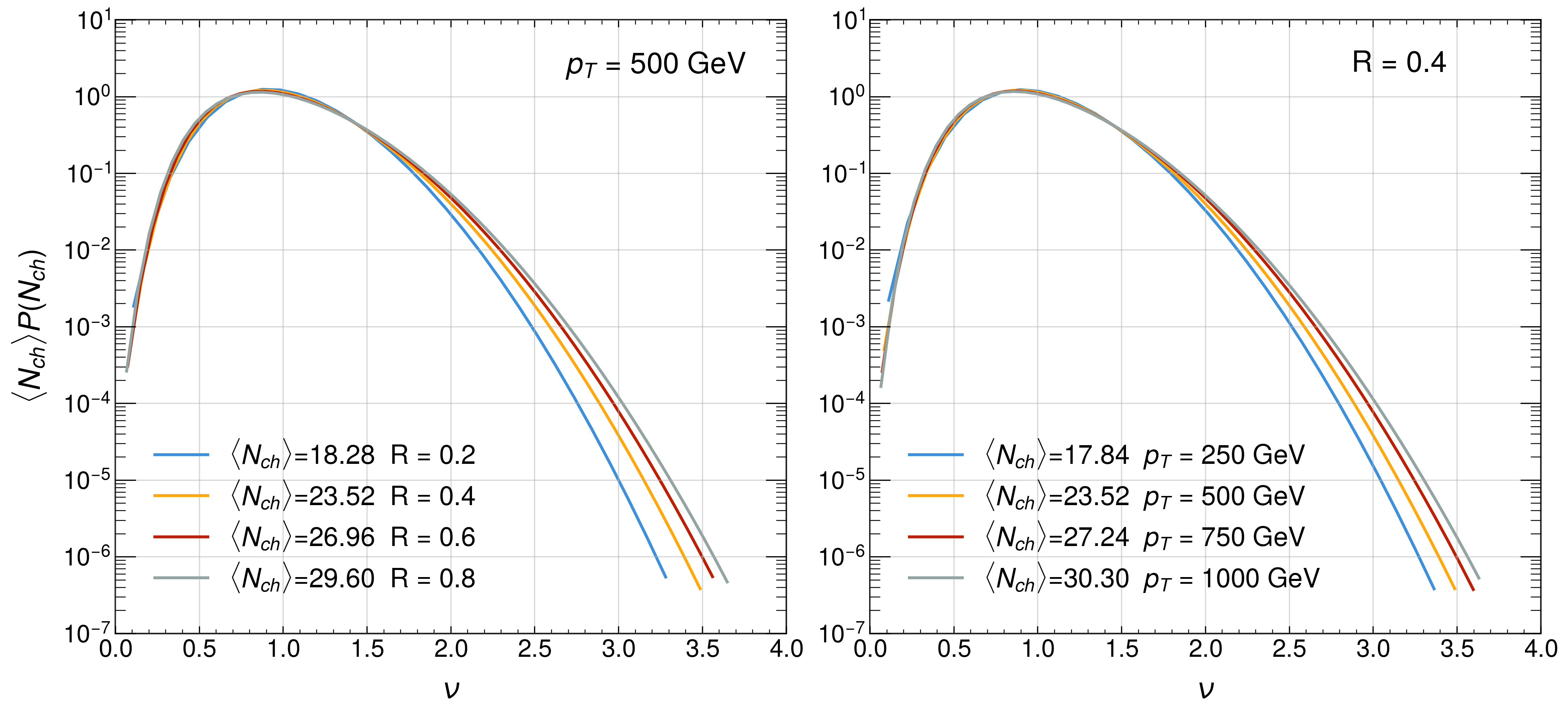}
    \caption{Rescaled charged-particle multiplicity distributions for exclusive gluon jets in the pure-gluon system, plotted as $\langle N_{\rm ch}\rangle P(\nu)$ versus $\nu = N_{\rm ch}/\langle N_{\rm ch}\rangle$. Left panel: distributions at fixed jet momentum $p_T=500$ GeV for jet radii $R=0.2, 0.4, 0.6, 0.8$. Right panel: distributions at fixed jet radius $R=0.4$ for $p_T=300, 500, 700, 900$ GeV. Mean multiplicities for each curve are indicated in the legend.}\label{fig:KNO-g2gg}
\end{figure}

The Koba-Nielsen-Olesen (KNO) scaling~\cite{Koba:1972ng,Lam:1983vw,Hinz:1985wq,Hegyi:2000sp,H1:2020zpd,Vertesi:2020utz,Liu:2022bru,Liu:2023eve,Germano:2024ier} states that multiplicity distributions $P(N)$ collapse onto a universal curve (denoted here as $\Psi(\nu)$) when expressed in terms of the normalized multiplicity,
\begin{align}
  \langle N_{ch} \rangle P_i(\nu \langle N_{\text{ch}}\rangle; \omega_J, R) = \Psi_i(\nu)\, .
\end{align}
regardless of the jet energy and cone size. Another way to state it is that the normalized distribution has all orders of cumulants independent of evolution. In fact, the double-logarithmic approximation (DLA) of the generating function does show KNO scaling. From Ref.~\cite{Bassetto:1983mvz}, taking the soft-gluon emission limit and approximating $p_{gg}(x) \approx 2C_A/[x(1-x)]$ and $\frac{1}{x}Z(s, x\omega_J)Z(s, (1-x)\omega_J)\approx \frac{1}{x}Z(s, \omega_J)Z(s, x\omega_J)$, the DLA generating function follows a second-order non-linear differential equation
\begin{align}
\frac{\mathrm{d}^2 \ln Z(s, t)}{\mathrm{d}t^2} &= e^{\ln Z(s,t)} - 1, \quad t = \sqrt{\frac{N_c\alpha_s}{2\pi} } \ln\frac{2\omega_J^2 \zeta}{Q_0^2}\,, \\
Z(s,0) &= \sum_n e^{-ns} P_0(n), \quad \left.\frac{\mathrm{d} Z(s,t)}{\mathrm{d} t}\right|_{t=0} = 0\,.
\end{align}
One can then obtains the cumulants equation by taking deriative with respect to $s$ according to Eq.~\eqref{eq:cumulants},
\begin{align}
\frac{\mathrm{d}^2 c_m}{\mathrm{d}t^2} = m! \sum_{\sum_{k'} k' j_{k'} = m}\prod_{k=1}^m \frac{c_k^{j_k}}{(k!)^{j_k} j_k!}\,.
\end{align}
Using the parton-hadron duality initial condition $P_0(n)=\delta_{n1}$, it allows a set of solutions $c_m(t) = \tilde{c}_m e^{mt}$, with $\tilde{c}_k$ determined order by order by the algebraic recursion relation
\begin{align}
m^2 \tilde{c}_m = m! \sum_{\sum_{k'} k' j_{k'} = m}\prod_{k=1}^m \frac{\tilde{c}_k^{j_k}}{(k!)^{j_k} j_k!}\,.
\end{align}
Recall that $c_1 = \langle n\rangle$, the resulting distribution indeed statisfies the requirement of the KNO scaling, i.e., $\frac{c_m(t)}{\langle n\rangle^m(t)} = \frac{\tilde{c_k}}{\tilde{c}_1^m}$. 

This also suggests that if one goes beyond DLA, there are several sources for the violation of the KNO scaling, including 1) a more realistic nonperturbative initial condition that includes fluctuations in the particle production, 2) using the full expression for the running coupling constant, 3) using the full splitting functions, and 4) using the exact energy fraction in non-linear term $Z(s, x\omega_J)Z(s, (1-x)\omega_J)$. Now, by numerically solving the full non-linear coupled evolution in Eqs.~\eqref{eq:zetaEvolution}, we can test the KNO scaling and the level of violation using the example of exclusive jets in a pure-gluon system. 

In Fig.~\ref{fig:KNO-g2gg}, the left panel shows $\langle N_{\rm ch}\rangle P_g(N_{\text{ch}})$ as a function of $\nu$ for fixed jet momentum ($p_T = 500~\mathrm{GeV}$) while varying the jet radius $R$, and the right panel shows the scaled distributions for fixed $R = 0.4$ while varying $p_T$. In both cases, there is an approximate KNO scaling for $\nu\lesssim 2$. For $\nu>2$, we observe a systematic breaking of the KNO scaling: with the increase of either $R$ or $p_T$, the large-$\nu$ tail of the normalized distributions broadens. The degree of violation gradually evolves with the characteristic jet scale $p_T R$. In sec.~\ref{sec:inclusiveJet}, we will use semi-inclusive jet function to perform the average over quark and gluon jet samples at the relevant collider energies, where we will also demonstrate that the breaking of the KNO scaling at large $\nu$ follows qualitatively the same trend as the \textsc{Pythia8} simulations at different jet energies and jet cone sizes.

\section{Internal multiplicity distribution of semi-inclusive jets}\label{sec:inclusiveJet}

\subsection{Semi-inclusive jet function with final-state flavor identification}\label{sec:RGE}
With the result for the exclusive jet case, we now calculate the semi-inclusive jet multiplicity function for both quark- and gluon-initiated jets. In contrast to the exclusive jet function, the semi-inclusive case introduces the variable $z$, which measures the fraction of the jet energy relative to that of the original parton. Moreover, the parton flavor is no longer fixed: the initiating quark or gluon can change flavor through the splitting process.

At LO, the jet is formed by a single parton carrying all its energy, which subsequently fragments into $n$ charged hadrons, so
\begin{align}
  \tilde{M}_i^{j, (0)}(s, z, \omega_J) &= J_{ji}^{(0)}(z,\omega_J) Z_j^{(0)}(s, \omega_J), \\
J_{ji}^{(0)}(z,\omega_J) &= \delta(1-z) \delta_{ij}
\end{align}
where $i$ denotes a quark ($q$) or a gluon ($g$).

\begin{figure}[ht]
    \centering
    \includegraphics[width=1.0\textwidth]{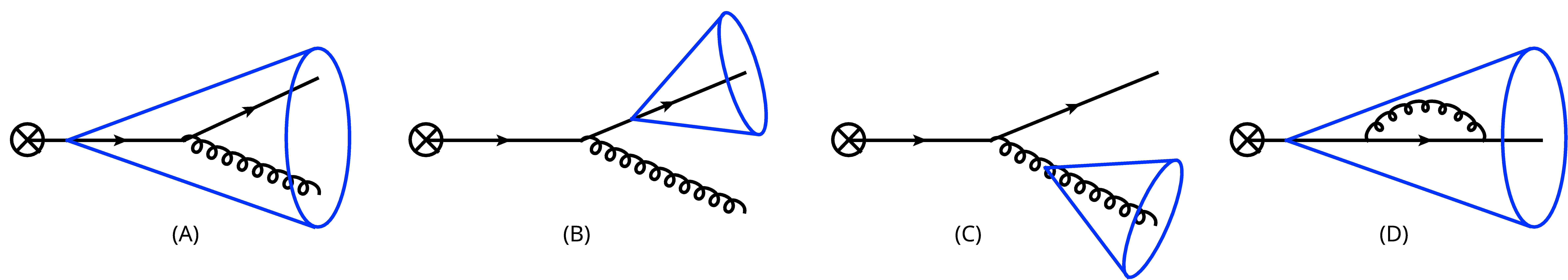}
    \caption{Four situations that contribute to the semi-inclusive quark jet multiplicity function in the light-cone gauge: (A) both quark and gluon are inside the jet, (B) only quark is inside the jet, (C) only gluon is inside the jet and (D) loop diagram.}\label{fig:quark_jet}
\end{figure}
At NLO, the calculation mostly follows the Ref.~\cite{Kang:2016mcy,Kang:2016ehg}, while the difference is that each parton in the final state is accompanied by a respective flavor-dependent multiplicity generating function. Therefore, the NLO correction is also classified according to the flavor of the final-state parton in the jet cone.
We continue to work in dimensional regularization with $d = 4 - 2\epsilon$ and perform the computation in light-cone gauge. As shown in Fig.~\ref{fig:quark_jet}, the semi-inclusive quark jet multiplicity function receives four contributions: (A), (B), (C), and (D). The loop diagram interfering with the LO diagram is scaleless in dimensional regularization, while the other three real-emission diagrams correspond to different kinematic configurations of parton emission within the anti-$k_T$ algorithm.

The treatment of soft radiation differs fundamentally depending on whether it remains inside the jet. For in-cone splittings (Configuration A), soft emissions are governed by QCD coherence and must be resummed using angular ordering, as implemented in Sec.~\ref{sec:angularOrdering}. For out-of-cone radiation (Configurations B and C), the dynamics simplify: soft partons outside the cone do not contribute to the jet multiplicity, and the finite energy of the observed jet ensures the initiating parton's energy fraction $z$ is bounded away from zero, regulating the $z \to 0$ divergence. This justifies using a standard virtuality-ordered evolution for out-of-cone radiation. Now we compute the contributions for all four cases.

\paragraph{Both partons contained in the jet} 
Fig.~\ref{fig:quark_jet} (A) shows the case where both partons remain inside the jet. Since the initiating quark transfers all of its energy to the jet, we have $z=\omega_J/\omega=1$. The corresponding one-loop bare semi-inclusive jet multiplicity function is therefore given by a delta function in $z$ times the exclusive jet multiplicity function obtained in Sec.~\ref{sec:exclusiveMulti}, so we have
\begin{align}\label{eq:quark1} 
\tilde{M}^{jk,(1)}_{i, \text{bare}}(s, z, \omega_J) =& \delta(1-z) \tilde{M}_{i, \text{bare}}^{jk,(1)}(s, \omega_J)\, .
\end{align}
The diagram in Fig.~\ref{fig:quark_jet} (D) interferes with the LO diagram. It vanishes because of the scaleless loop integral. It will only alter the nature of the IR divergence to UV for the piece of the semi-inclusive jet function that is proportional to $\delta(1-z)$.
\paragraph{Single parton contained in the jet}  
Fig.~\ref{fig:quark_jet} (B) and (C) show the case where only one daughter parton from the parent quark's splitting is contained within the jet. The jet is therefore formed by either the quark or the gluon, carrying a fraction $\omega_J = z\omega$ of the initial energy. For a generic splitting $i \to j k$, the parton $j$ that remains inside the jet determines the jet flavor and the subsequent multiplicity evolution within the jet:
\begin{align}
\tilde{M}^{j(k),(1)}_{i,\text{bare}}(s, z, \omega_J) 
&= \frac{\alpha_s}{\pi} \frac{{(\mu^2 e^{\gamma_E})}^{\epsilon}}{\Gamma(1-\epsilon)} \hat{p}_{ji}(z, \epsilon) Z_j^{(0)}(s, \omega_J)  \int \frac{\mathrm{d} k_{\perp}}{k_{\perp}^{1+2\epsilon}} \Theta_{\text{alg}}^{>R}, \label{eq:quark2}
\end{align}
Here, `$j(k)$' indicates that parton $j$ is inside the jet and parton $k$ is outside. The jet algorithm imposes the constraint that one of the partons lies outside the jet. For the anti-$k_T$ algorithm, this condition is expressed as~\cite{Kang:2016mcy}
\begin{align}
\Theta_{\text{alg}}^{> R} 
   = \theta {\left(k_{\perp} - (1-z) \omega_J \tan \frac{\mathcal{R}}{2}\right)}, \label{eq:antiOut}
\end{align}
Which leads to the $k_{\perp}$ integral as
\begin{align}
\int_0^\infty \frac{\mathrm{d} k_{\perp}}{k_{\perp}^{1+2\epsilon}} \Theta_{\text{alg}}^{> R}  = \frac{1}{2\epsilon} {\left( \omega_J \tan \frac{\mathcal{R}}{2} \right)}^{-2\epsilon} {\left( 1-z \right)}^{-2\epsilon}\,. \label{eq:antiInteOut}
\end{align}
If there is only one particle $j(k)$ contained in the cone, then the NLO correction to the multiplicity generating function is linearly proportional to $Z_j(k)^{(0)}$ with the same energy of the jet. All the non-trivial corrections to $Z$ are due to the flavor changing $i\rightarrow j(k)$ of the parton in the jet.
With such an understanding, we give the NLO correction with only one parton inside the jet cone
\begin{align}
     \tilde{M}^{q(g),(1)}_{q,\text{bare}}(s, z, \omega_J) 
    &= \frac{\alpha_s}{2\pi}\left\{\delta(1-z) C_F {\left[ -\frac{1}{{\epsilon}^2} - \frac{1}{\epsilon}\mathcal{L} - \frac{1}{2}\mathcal{L}^2 -\frac{3}{2 \epsilon} - \frac{3}{2}\mathcal{L} + \frac{\pi^2}{12} \right]} \right. \nonumber \\
    & \left. + \left( \frac{1}{\epsilon} + \mathcal{L} \right) \hat{P}_{qq}(z)  - C_F ( 1-z ) - 2 C_F ( 1 + z^2) {\left[\frac{\ln (1-z)}{1-z} \right]}_+ \right\} Z_q^{(0)}(s, \omega_J), \label{eq:qToq} \\
     \tilde{M}^{(q)g, (1)}_{q,\text{bare}}(s, z, \omega_J) 
    &=  \frac{\alpha_s}{2\pi}\left\{ {\left( \frac{1}{\epsilon} + \mathcal{L} \right)} \hat{P}_{gq}(z) -  \hat{P}_{gq}(z) 2 \ln(1-z) - C_F z  \right\}Z^{(0)}_g(s, \omega_J), \\
    \tilde{M}_{g,\text{bare}}^{g(g)(1)}(s, z, \omega_J)
     &=  \frac{\alpha_s}{2\pi} \left\{\delta(1-z)\left[ -\frac{C_A}{{\epsilon}^2} - \frac{C_A}{\epsilon}\mathcal{L} - \frac{C_A}{2} \mathcal{L}^2 - \frac{\beta_0}{2 \epsilon} - \frac{\beta_0}{2} \mathcal{L} + \frac{\pi^2}{12}C_A \right] \right. \nonumber \\
     &\left.  + {\left( \frac{1}{\epsilon} + \mathcal{L} \right)} \hat{P}_{gg}(z)   - \frac{4C_A {[1-z(1-z)]}^2}{z}  {\left[ \frac{\ln (1-z)}{1-z} \right]}_+ \right\} Z_g^{(0)}(s, \omega_J), \\
    \tilde{M}_{g,\text{bare}}^{q(\bar{q}),(1)}(s, z, \omega_J)
    &=\frac{\alpha_s}{2\pi}  2n_f \left\{{\left( \frac{1}{\epsilon} + \mathcal{L} \right)}  \hat{P}_{qg}(z) \right.\nonumber\\
    & \hskip2cm \left.- 2 \hat{P}_{qg}(z) \ln(1-z) - T_F 2z(1-z)  \right\} Z^{(0)}_q(s, \omega_J).
\end{align}
with the logarithm $\mathcal{L} = \ln \frac{\mu^2}{\omega_J^2 \tan^2 \frac{\mathcal{R}}{2}}$. Note that all the corrections are for the semi-inclusive jet function, but not for the multiplicity generating function inside the jet ($Z_q^{(0)}$ factors out).

With the above expressions, we obtain the following result for $\mathcal{O}(\alpha_s)$ correction to the bare semi-inclusive jet multiplicity function as
\begin{align}
  \tilde{M}_{i,\text{bare}}^j(s, z, \omega_J, \mu) &= \tilde{M}_{i, \text{bare}}^{j,(0)}(s, z, \omega_J, \mu)\delta_{ij} + \tilde{M}_{i, \text{bare}}^{j k, (1)}(s, z, \omega_J, \mu) + \tilde{M}_{i, \text{bare}}^{j (k), (1)}(s, z, \omega_J, \mu), \nonumber\\
  &= J_{ji} Z_j + \mathcal{O}(\alpha_s^2) \,.
\end{align}
Note that the LO results need only be added for the case $i=j$. We obtain the following results up to $\mathcal{O}(\alpha_s)$ for the semi-inclusive jet function and the multiplicity generating function. The results for the $Z$ functions, its NLO corrections and renormalization are precisely the same as the case for the exclusive jet multiplicity generating function as in Sec.~\ref{sec:renormalization}, since we have seen that the radiation out of the cone does not introduce extra radiative correction to $Z$. 

In the final result, only single poles $1/\epsilon$ and single logarithmic terms $\mathcal{L}$ survive. All double poles $1/\epsilon^2$ and double logarithms $\mathcal{L}^2$ that arise during intermediate steps cancel out for semi-inclusive jet functions, as in Ref.~\cite{Kang:2016mcy}. The difference is that different jet flavors must be multiplied by their respective multiplicity generating function, what we need is not the standard jet function $J_q$ and $J_g$, but the transfer matrix $J_{ji}$. Here we list the results accurate to NLO. They are almost the same as the standard semi-inclusive jet function, but with identification of the final parton
\begin{align}
  J_{qq}^{\text{bare}}(z, \omega_J, \mu) 
&= \delta(1-z)\left[1+\frac{\alpha_s}{2\pi}  \left(d^{q}_{\textrm{anti-$k_T$}}+C_F\frac{\pi^2}{12} \right) \right] \nonumber\\
& +  \frac{\alpha_s}{2\pi}\left( \frac{1}{\epsilon} + \mathcal{L} \right) \hat{P}_{qq}(z)  - \frac{\alpha_s}{2\pi} C_F \left[  (1-z) + 2(1 + z^2) {\left[\frac{\ln (1-z)}{1-z} \right]}_+ \right]  \,,\\
J_{gq}^{\text{bare}}(z, \omega_J, \mu) 
&= \frac{\alpha_s}{2\pi}\left( \frac{1}{\epsilon} + \mathcal{L} \right)\hat{P}_{gq}(z) - \frac{\alpha_s}{2\pi}\left[ \hat{P}_{gq}(z) 2 \ln(1-z) + C_F z \right] \,,\\
J_{gg}^{\text{bare}}(z, \omega_J, \mu) 
& = \delta(1-z)\left[1 + \frac{\alpha_s}{2\pi} \left(d^{g}_{\textrm{anti-$k_T$}}+C_A\frac{\pi^2}{12} \right) \right] \nonumber\\
& + \frac{\alpha_s}{2\pi}\left( \frac{1}{\epsilon} + \mathcal{L} \right) \hat{P}_{gg}(z)  - \frac{\alpha_s}{2\pi}\frac{4C_A {\left( 1-z(1-z) \right)}^2}{z} {\left[ \frac{\ln (1-z)}{1-z} \right]}_+ \,,\\
J_{qg}^{\text{bare}}(z,\omega_J, \mu) 
&= \frac{\alpha_s}{2\pi} 2n_f \left\{\left( \frac{1}{\epsilon} + \mathcal{L} \right)  \hat{P}_{qg}(z)  -  2\left[ \hat{P}_{qg}(z) \ln(1-z) + T_F z(1-z) \right] \right\}\,.
\end{align}

\subsection{Renormalization and evolution of the flavor-identified SiJF}
The renormalization of the multiplicity generating function proceeds exactly as in the exclusive case, as discussed in Sec.~\ref{sec:renormalization}. For the semi-inclusive case, one additionally needs the renormalization of the semi-inclusive jet function (SiJF). Since the structure of SiJF renormalization is already well established in the literature and shown to follow DGLAP evolution~\cite{Kang:2016mcy}, we do not repeat the full derivation and only summarize the relevant results and incorporate final-state flavor identification. The renormalized SiJF with identified final flavor $j=q,g$ then obeys the coupled DGLAP-type evolution equations
\begin{align}\label{eq:SIJFDGLAP} 
\frac{\partial}{\partial \ln \mu^2} 
\begin{bmatrix}
J_{jq}(z,\omega_J, \mu)\\
J_{jg}(z,\omega_J, \mu)
\end{bmatrix}
= \frac{\alpha_s(\mu^2)}{2 \pi}
\begin{bmatrix}
\hat{P}_{qq}(z) & \hat{P}_{gq}(z)\\
2n_f \hat{P}_{qg}(z) & \hat{P}_{gg}(z)
\end{bmatrix}
 \otimes_z
 \begin{bmatrix}
J_{jq}(z,\omega_J, \mu)\\
J_{jg}(z,\omega_J, \mu)
\end{bmatrix}\,.
\end{align}
The symbol $\otimes_z$ denotes the convolution in momentum fraction,
\begin{align}
  f \otimes_z g = \int_{z}^{1} \frac{\mathrm{d} z'}{z'} f\left(\frac{z}{z'}\right) g(z')\, .
\end{align}
It is convenient to solve Eq.~\eqref{eq:SIJFDGLAP} in the Mellin moment space. The Mellin transformation of function $f(z)$ is defined by
\begin{align}
  f(N) = \int_{0}^{1} \mathrm{d}z\, z^{N-1} f(z)\,, \qquad
  (f \otimes_z g)(N) = f(N) g(N)\,.
\end{align}
The solution of the transformed Eq.~\eqref{eq:SIJFDGLAP} is then obtained from the standard procedures in Ref.~\cite{Vogt:2004ns}. Diagonalizing the matrix of the LO splitting functions in Eq.~(\ref{eq:SIJFDGLAP}) in the moment space
\begin{align}
\begin{bmatrix}
\hat{P}_{qq}(N) & \hat{P}_{gq}(N)\\
2n_f \hat{P}_{qg}(N) & \hat{P}_{gg}(N)
\end{bmatrix} = U_N^{-1}\mathrm{diag}\{\lambda_+(N),\,\lambda_-(N)\} U_N
\end{align}
where, $\lambda_{\pm}$ are the two eigenvalues and $U_N$ is the matrix for similarity transformation.
The solution to $\mathbf{X}_N(\mu) = \left[J_{jq}\left(N,\omega_J, \mu\right),  J_{jg}\left(N,\omega_J, \mu\right)\right]$ is compactly written as 
\begin{align}
\mathbf{X}_N^T(\mu)=
U_N^{-1} \mathrm{diag}\left\{
\left(\frac{\alpha_s(\mu_J^2)}{\alpha_s(\mu^2)}\right)^{\frac{2}{\beta_0}\lambda_+(N)},\, 
\left(\frac{\alpha_s(\mu_J^2)}{\alpha_s(\mu^2)}\right)^{\frac{2}{\beta_0}\lambda_-(N)}
\right\} U_N 
\mathbf{X}_N^T(\mu_J)\,.
\end{align}
The solution in $z$-space is then obtained by inverse Mellin transformation,
\begin{align}
  f(z)=\frac{1}{2\pi i}\int_{\mathcal{C}_N} \mathrm{d}N\, z^{-N} f(N),
\end{align}
with the contour $\mathcal{C}_N$ chosen to the right of all singularities also shown in Ref.~\cite{Vogt:2004ns}.

The equations separate into two groups depending on whether we want to identify a quark or gluon jet that initiated the subsequent multiplicity evolution. Although the evolution kernels are identical, the initial conditions differ. For a jet with energy $\omega_J$ and radius $\mathcal{R}$, we choose the natural starting scale as $\mu_J \equiv \omega_J \tan\!\frac{\mathcal{R}}{2} \approx p_T R$. At this scale, the logarithmically enhanced contributions vanish, and the renormalized NLO initial condition for the semi-inclusive jet function is given by
\begin{equation}\label{eq:initialJet}
  J_{ji}(z,\omega_J, \mu_J) = J_{ji}^{(0)}(z,\omega_J, \mu_J) +  J_{ji}^{(1)}(z,\omega_J, \mu_J).
\end{equation}
The LO order expression is given by 
\begin{align}
\label{eq:SiJF-IC-LO}
J_{ji}^{(0)}(z,\omega_J,\mu_J) = \delta_{ji}\delta(1-z).
\end{align}
The NLO expressions are
\begin{subequations}
\label{eq:SiJF-IC-NLO}
\begin{align}
J_{qq}^{(1)}(z, \omega_J,\mu_J) 
&= \frac{\alpha_s(\mu_J^2)}{2\pi}\left\{\delta(1-z) \left(d^{q}_{\textrm{anti-$k_T$}} + C_F\frac{\pi^2}{12} \right)\right.\nonumber\\
&  \quad \left. - 2C_F(1 + z^2){\left[\frac{\ln (1-z)}{1-z} \right]}_+ - C_F (1-z)\right\}\,,\\
J_{gq}^{(1)}(z, \omega_J,\mu_J) 
&= \frac{\alpha_s(\mu_J^2)}{2\pi}\left\{ -  \hat{P}_{gq}(z) 2 \ln(1-z) - C_F z \right\}\,, \\
J_{gg}^{(1)}(z, \omega_J,\mu_J) 
& = \frac{\alpha_s(\mu_J^2)}{2\pi}\left\{\delta(1-z)\left(d^{g}_{\textrm{anti-$k_T$}}+C_A\frac{\pi^2}{12} \right) \right. \nonumber\\
&\quad \left.- \frac{4C_A {\left( 1-z(1-z) \right)}^2}{z} \left[ \frac{\ln (1-z)}{1-z} \right]_+ \right\}\,, \\
J_{qg}^{(1)}(z, \omega_J,\mu_J) 
&= \frac{\alpha_s(\mu_J^2)}{2\pi}2n_f\left\{ -2\hat{P}_{qg}(z) \ln(1-z) - T_F 2z(1-z)  \right\}\, .
\end{align}
\end{subequations}
These expressions provide the flavor-resolved initial conditions that are subsequently evolved to higher scales via the coupled DGLAP equations, thus enabling a consistent treatment of semi-inclusive jet multiplicity functions with explicit flavor tagging.

\subsection{Inclusive jet cross section with multiplicity selection}\label{sec:factorizationFormula}

We now present the factorized expression for the semi-inclusive jet multiplicity function with multiplicity selection. The cross section for producing a jet of radius $R$ with transverse momentum $p_{T,J}$ and containing exactly $n$ charged particles is given by:
\begin{align}\label{eq:factorizationTwoScale}
  \frac{\mathrm{d} \sigma_{pp \to J(n)+X}}{\mathrm{d} p_{T,J} \mathrm{d} \phi_J \mathrm{d} \eta_J}
&= \sum_{ij} \bigg\{\mathrm{d} \sigma_{pp\to i}^{(0)}(p_{T,J}/z,\mu_H)\otimes_z J_{ji}^{(0)}(z,\mu_H,\mu_J) \times P_{j}^{(0)}(n,\zeta_R,\zeta_0) \nonumber \\
&\quad +\mathrm{d} \sigma_{pp \to i}^{(1)}(p_{T,J}/z,\mu_H)\otimes_z J_{ji}^{(0)}(z,\mu_H,\mu_J) \times P_{j}^{(0)}(n,\zeta_R,\zeta_0) \nonumber \\
&\quad +\mathrm{d} \sigma_{pp \to i}^{(0)}(p_{T,J}/z,\mu_H)\otimes_z J_{ji}^{(1)}(z,\mu_H,\mu_J) \times P_{j}^{(0)}(n,\zeta_R,\zeta_0) \nonumber \\
&\quad +\mathrm{d} \sigma_{pp \to i}^{(0)}(p_{T,J}/z,\mu_H)\otimes_z J_{ji}^{(0)}(z,\mu_H,\mu_J) \times P_{j}^{(1)}(n,\zeta_R,\zeta_0) \bigg\}.
\end{align}
$\mathrm{d}\sigma_{pp\rightarrow i}^{(0)}$ and $\mathrm{d}\sigma_{pp\rightarrow i}^{(1)}$ are the $\mathcal{O}(\alpha_s^2)$ and $\mathcal{O}(\alpha_s^3)$ parts of the hard partonic cross sections~\cite{AVERSA1989105,Jager:2002xm} at hard scale $\mu_H=p_{T,J}/z$.  
The semi-inclusive jet function $J_{ji}(z, \mu_H, \mu_J)$ is obtained by first initializing it at its natural scale $\mu_J = p_T R$, where its fixed-order expression contains no large logarithms as shown in Eq.~\eqref{eq:initialJet}. Then, perform the DGLAP evolution from $\mu_J$ to the hard scale $\mu_H$. Finally, convolved with the partonic cross sections to compute the jet production cross section, differentially in the jet flavor $j$, the divergence around $z \to 1$ is carefully handled with the strategy described in~\cite{Bodwin:2015iua,Kang:2016mcy}.
$J_{ji}^{(0)}$ and $J_{ji}^{(1)}$ are calculated with the leading-order and the next-to-leading order jet function in Eqs.~(\ref{eq:SiJF-IC-LO}) and (\ref{eq:SiJF-IC-NLO}), respectively.
The multiplicity distribution function $P_j(n, \zeta_R, \zeta_0)$ is obtained by the following procedure. First, we specify a nonperturbative model for multiplicity generating function $Z_j(s, \omega_J, \zeta_0)$ at a small angle $\zeta_0 = 1 - \cos \theta_0$ as discussed in Sec~\ref{sec:NPModel}.
 Then, we evolve the multiplicity generating function using the angular-ordered coupled non-linear equations (Eqs.~\eqref{eq:zetaEvolution}) from $\zeta_0$ to the jet boundary $\zeta_R$. In the end, the generating function is inverted to yield the final probability distribution $P_j(n)$ for a jet initiated by a parton $j$ with Eq.~\eqref{eq:invertLaplace}. Here, $P_{j}^{(0)}$ is the multiplicity distribution function inverted from the evolved nonperturbative initial conditions in Eqs.~(\ref{eq:NP-IC-Z}) and (\ref{eq:NP-IC-Navg}); 
$P_{j}^{(1)}$ is the NLO leftover terms inverted from Eqs.~(\ref{eq:NLOcoefficientzeta}). 

The per-jet normalized multiplicity distribution, which is the primary observable for comparison with experimental data, is defined as:
\begin{align}
P_J(n) = {\left[\frac{\mathrm{d}\sigma_{pp\rightarrow J+X}}{\mathrm{d}p_{T, J} \mathrm{d}\phi_J \mathrm{d}\eta_J}\right]}^{-1}\times  \frac{\mathrm{d}\sigma_{pp\rightarrow J(n)+X}}{\mathrm{d}p_{T,J} \mathrm{d}\phi_J \mathrm{d}\eta_J}\,,
\end{align}
with the corresponding semi-inclusive jet cross section (without multiplicity selection) given by
\begin{align}
\label{eq:inclusiveJetCrossSection}
\frac{\mathrm{d} \sigma_{pp\rightarrow J+X}}{\mathrm{d} p_{T,J} \mathrm{d} \phi_J \mathrm{d} \eta_J}
&= \sum_{ij} \bigg\{\mathrm{d} \sigma_{pp\rightarrow i}^{(0)}(p_{T,J}/z,\mu_H)\otimes_z J_{ji}^{(0)}(z,\mu_H,\mu_J) \nonumber\\
&\quad + \mathrm{d} \sigma_{pp\rightarrow i}^{(1)}(p_{T,J}/z,\mu_H)\otimes_z J_{ji}^{(0)}(z,\mu_H,\mu_J) \nonumber \\
&\quad + \mathrm{d} \sigma_{pp\rightarrow i}^{(0)}(p_{T,J}/z,\mu_H)\otimes_z J_{ji}^{(1)}(z,\mu_H,\mu_J) \bigg\}\,.
\end{align}
Only combinations of $\mathrm{d}\sigma_{pp\rightarrow i}$, $J_{ji}$ and $P_j$ up to order $\alpha_s^3$ correction is retained in Eqs.~(\ref{eq:factorizationTwoScale}) and~(\ref{eq:inclusiveJetCrossSection}).
This defines a probability distribution ($\sum_n P_J(n) = 1$) for the number of particles inside a jet of given $p_T$ and $R$. To test the effect of NLO corrections in $\alpha_s$, we define two levels of perturbative accuracy by
\begin{itemize}
\item LO+LL: LO hard partonic cross sections, convolved with LO jet function and multiplicity distribution function. Both the jet function and the multiplicity distribution are improved with RG equations: LL$_R$ DGLAP resummation for jet functions, and the angular-ordered coupled nonlinear equations with leading-order QCD splitting function for the generating functions.
\item NLO+LL: both the hard partonic cross sections, jet function, and multiplicity distribution function are accurate to NLO. The RG equations are the same as in the first case.
\end{itemize}

\begin{figure}[!t]
    \centering
    \includegraphics[width=1.0\linewidth]{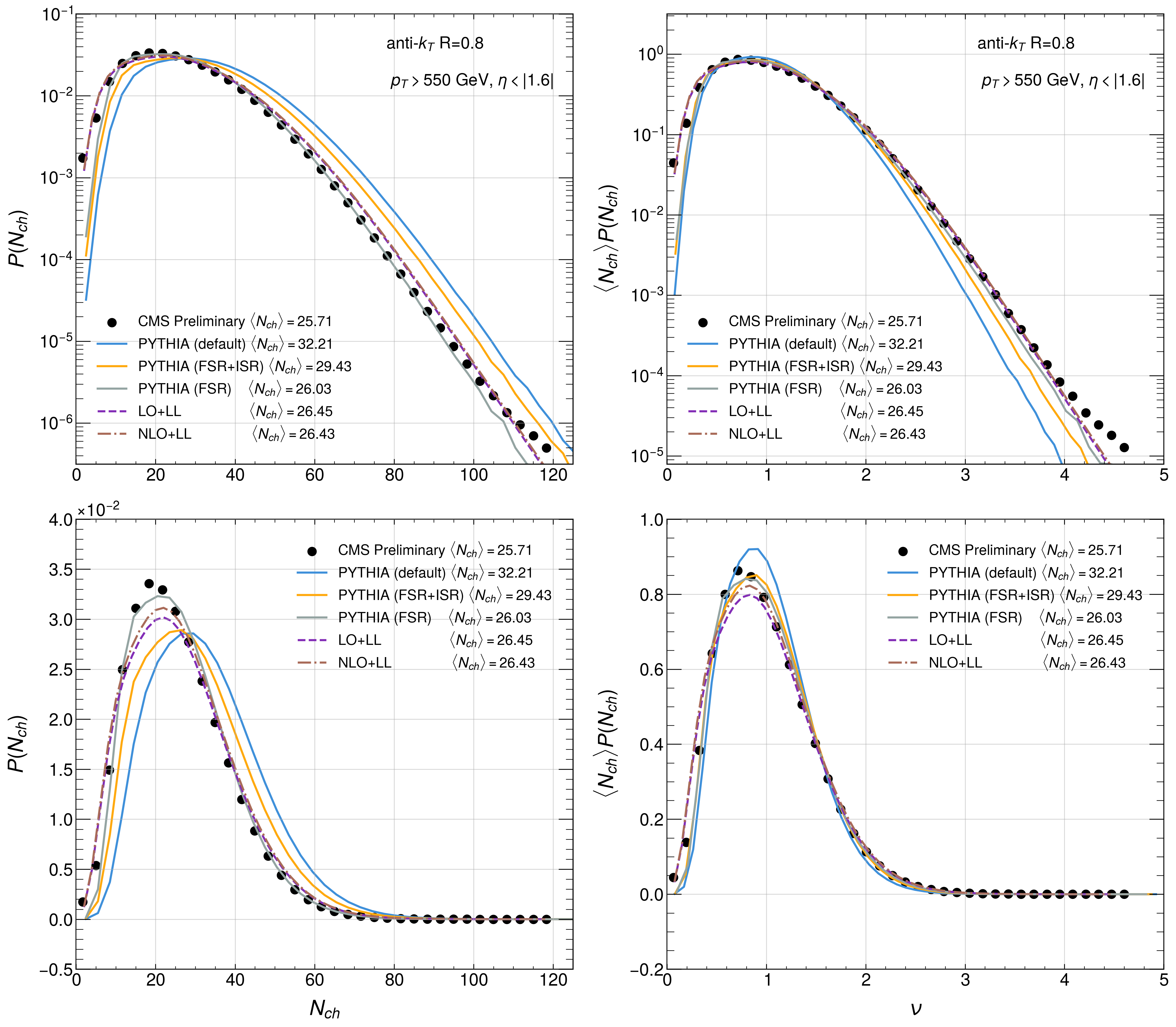}
    \caption{Charged-particle multiplicity distributions $P(N_{\text{ch}})$ (left) and rescaled distribution $\langle N_{\text{ch}}\rangle P(N_{\text{ch}})$ as a function of $\nu = N_{\mathrm{ch}} / \langle N_{\mathrm{ch}} \rangle$ (right). Top row shows the distributions in logarithmic scale, bottom row in linear scale, mean multiplicity is explicitly labeled inside each panel..  We compare our LO+LL (purple dashed) and NLO+LL (brown dash-dot) calculations, \textsc{Pythia8} (default) (blue), \textsc{Pythia8} (FSR+ISR) (orange), \textsc{Pythia8} (FSR) (grey) and CMS data (black points) for jets with $p_T > 550~\mathrm{GeV}$ and $R=0.8$ and $\eta < |1.6|$. The data are taken from the datasets of Refs.~\cite{CMS:2023iam, CMS_HIN_21_013}.}\label{fig:compareCMS}
\end{figure}

\section{Results for internal multiplicity distribution of semi-inclusive jets}\label{sec:resultsSemiInclusive}
\paragraph{Comparison to CMS data} We use CMS measurements of charged-particle multiplicity inside jets in $pp$ collisions at $\sqrt{s}=13~\mathrm{TeV}$ with $p_T > 550~\mathrm{GeV}$ and jet radius $R=0.8$ as our primary experimental benchmark~\cite{CMS:2023iam}. 
The main theoretical calculations at LO+LL and NLO+LL order are already described in the previous section.

In order to estimate the impact of other physical processes in realistic proton-proton collisions, we also compare to  \textsc{Pythia8}~\cite{Bierlich:2022pfr} simulations in three different setups: (i) the default tune, including initial-state radiation (ISR), multi-parton interactions (MPI), final-state radiation (FSR); (ii) turn off MPI with only ISR, FSR enabled, and (iii) only FSR enabled.
The MC jets are reconstructed using the anti-$k_T$ algorithm, as implemented in \textsc{FastJet}~\cite{Cacciari:2008gp,Cacciari:2011ma}, with the same kinematic selections as in the CMS analysis. 
After clustering all stable particles into jets, the charged constituents are identified to obtain the charged-particle multiplicity inside each jet.
Here, case (iii) provides the most direct connection with our theory calculation, where only the primary hard scatterings and parton shower effects in the final state are included in shaping the multiplicity distribution. 

Fig.~\ref{fig:compareCMS} compares the CMS data (symbols, not yet unfolded)~\cite{CMS:2023iam, CMS_HIN_21_013}, the theoretical calculations at two accuracies (dashed and dash-dotted lines), and the three cases of MC simulations (solid lines). The top left panel shows the original distribution, while the top right panel compares the normalized distribution.
The bottom row contains essentially the same information but plotted on a linear scale to resolve details near the peak of the distribution.

Focusing on the left column of Fig.~\ref{fig:compareCMS}, we find that both the LO+LL and NLO+LL calculations provide a good description of the CMS data. The two curves are nearly indistinguishable on the logarithmic scale, while on the linear scale one can observe that the NLO correction slightly enhances the peak height relative to the LO result. 

The case (iii) \textsc{Pythia8} simulation with only FSR enabled agrees well with our theoretical prediction, as both incorporate essentially the same perturbative dynamics of the final-state parton shower. One noticeable feature is that the \textsc{Pythia8} result systematically undershoots the data in the low-multiplicity region, $N_{\rm ch}<5\text{--}10$, whereas our theoretical calculation, supplemented with a simple nonperturbative parametrization, provides a better description there. When ISR and MPI are included in the \textsc{Pythia8} simulations, uncorrelated particles from the underlying event contaminate the multiplicity distribution inside the jet cone, leading to an overall shift of the spectrum toward larger multiplicities.

\begin{figure}[!b]
    \centering
    \includegraphics[width=1.0\textwidth]{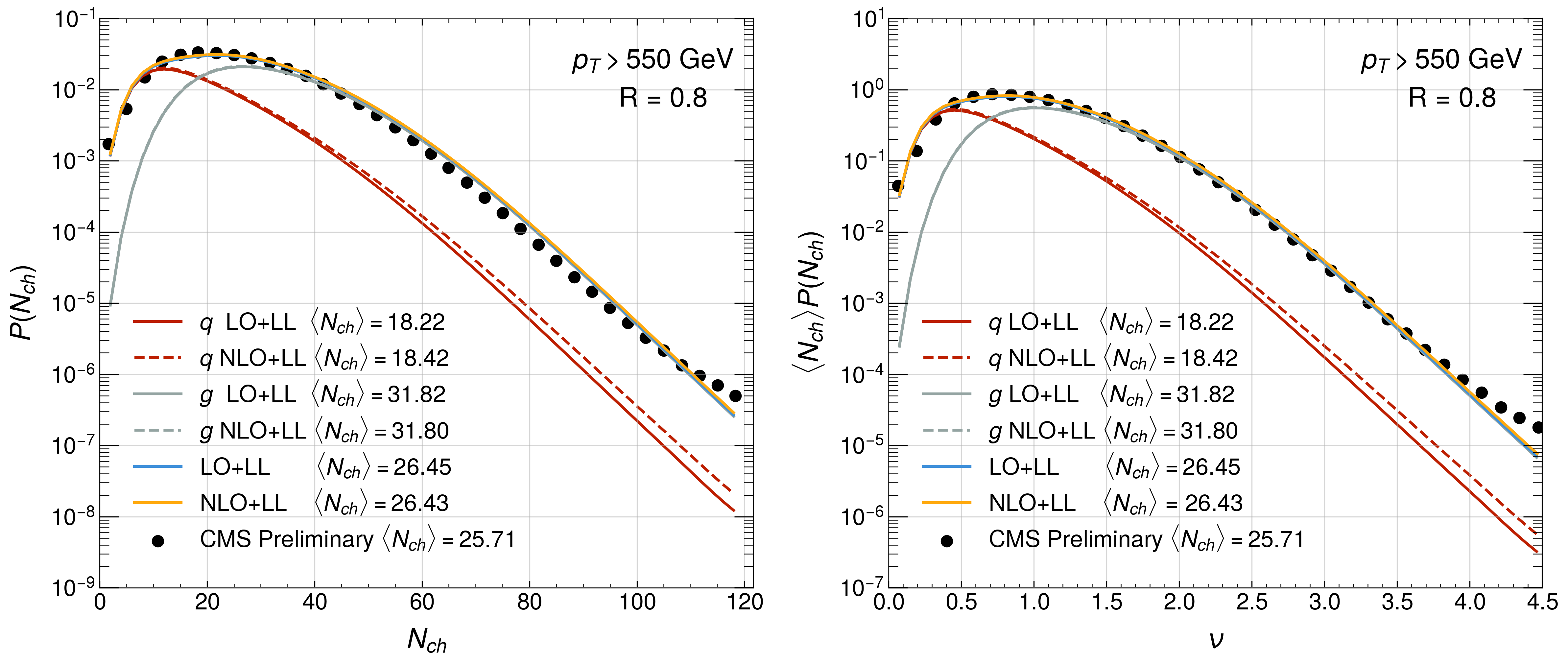}
    \caption{Charged-particle multiplicity distributions $P(N_{\text{ch}})$ (left) and rescaled distributions $\langle N_{\rm ch}\rangle P(N_{\text{ch}})$ as a function of $\nu$ (right) for jets with $R=0.8$ and $p_T>550\,\mathrm{GeV}$. Results from LO+LL evolution (blue) and NLO+LL evolution (orange) are compared with CMS data (black dots). Also shown are the quark (red) and gluon (gray) contributions at LO+LL (solid) and NLO+LL (dashed), each weighted by their relative semi-inclusive jet cross-section fractions so that they sum to the semi-inclusive result.}\label{fig:Pn_q_vs_g_R0_8}
\end{figure}

Turning to the right column of Fig.~\ref{fig:compareCMS}, we find that after rescaling by the mean multiplicity, both our calculation and the \textsc{Pythia8} (FSR) simulation reproduce the overall shape of the distribution quite well, but systematically underestimate the very high-multiplicity tail ($\nu \gtrsim 4$). At $\nu = 4.5$, the discrepancy reaches nearly a factor of three. As for the \textsc{Pythia8} simulation including ISR and MPI effects, the additional soft activity increases the mean multiplicity $\langle N_{\rm ch}\rangle$, thereby narrowing the normalized distribution and leading to a significant deviation from the CMS data.

As discussed in the context of nonperturbative (NP) effects, any tuning of the NP model typically modifies both the low-multiplicity region ($\nu \ll 1$) and the high-multiplicity tail ($\nu \gg 1$) simultaneously. Therefore, the persistent deficit at $\nu > 4$ is unlikely to be resolved by NP adjustments alone and is more plausibly of perturbative origin. One possible improvement would be to incorporate higher-order nonlinear effects in the evolution, such as intrinsic $1\!\to\!3$ or more general $1\!\to\!n$ splitting kernels, which we plan to explore in a forthcoming study.

\begin{figure}
    \centering
    \includegraphics[width=1.0\textwidth]{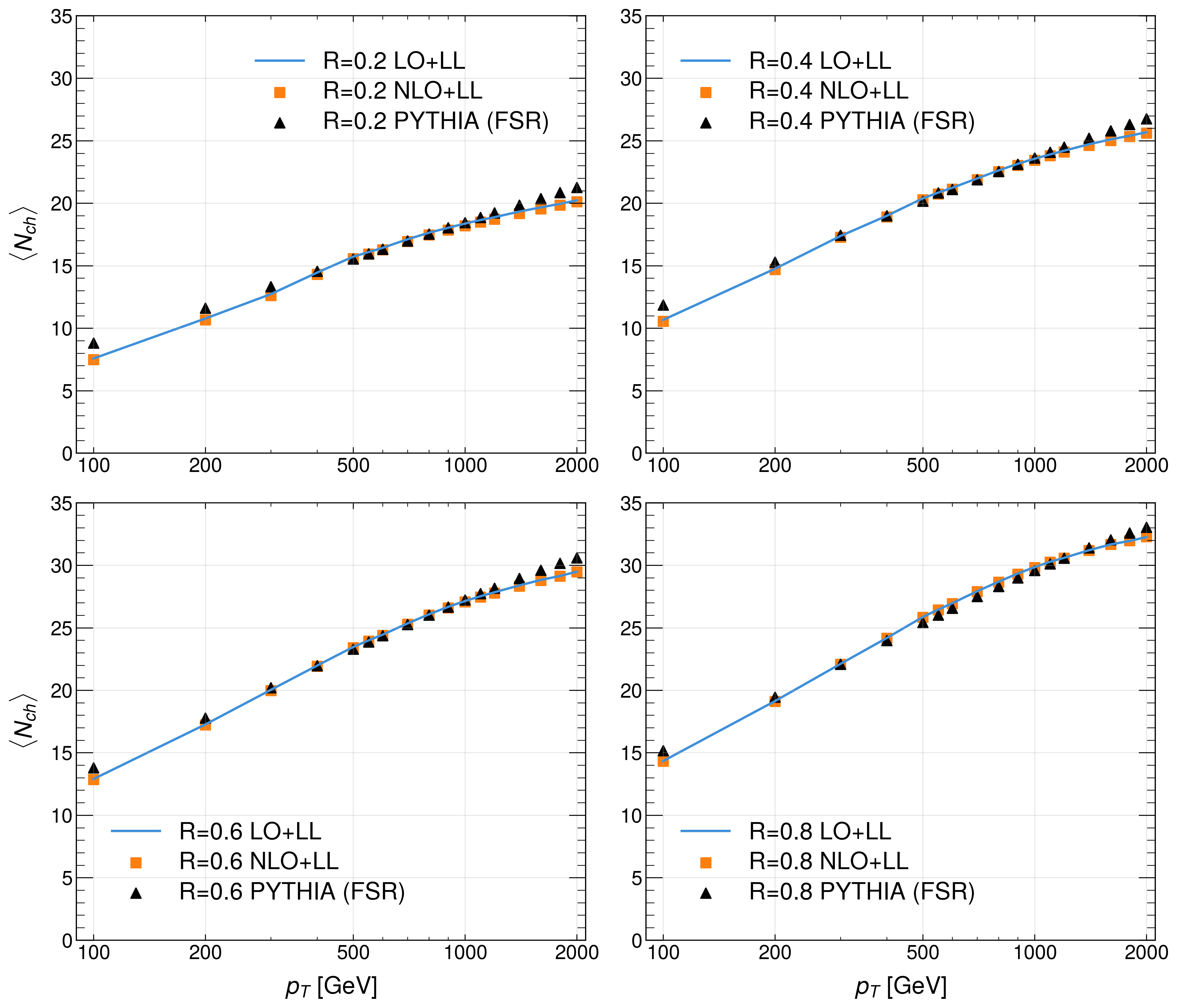}    
    \caption{Mean charged particle multiplicity $\langle N_{\rm ch} \rangle$ as a function of jet transverse momentum $p_T$ for jets with different radii $R = 0.2, 0.4, 0.6, 0.8$. 
    LO+LL predictions are shown as blue solid lines, NLO+LL predictions as orange solid squares, and \textsc{Pythia8} simulations (FSR only) as black solid triangles.
    }\label{fig:Nmean_Rs}
\end{figure}
\begin{figure}
    \centering
    \includegraphics[width=1.0\textwidth]{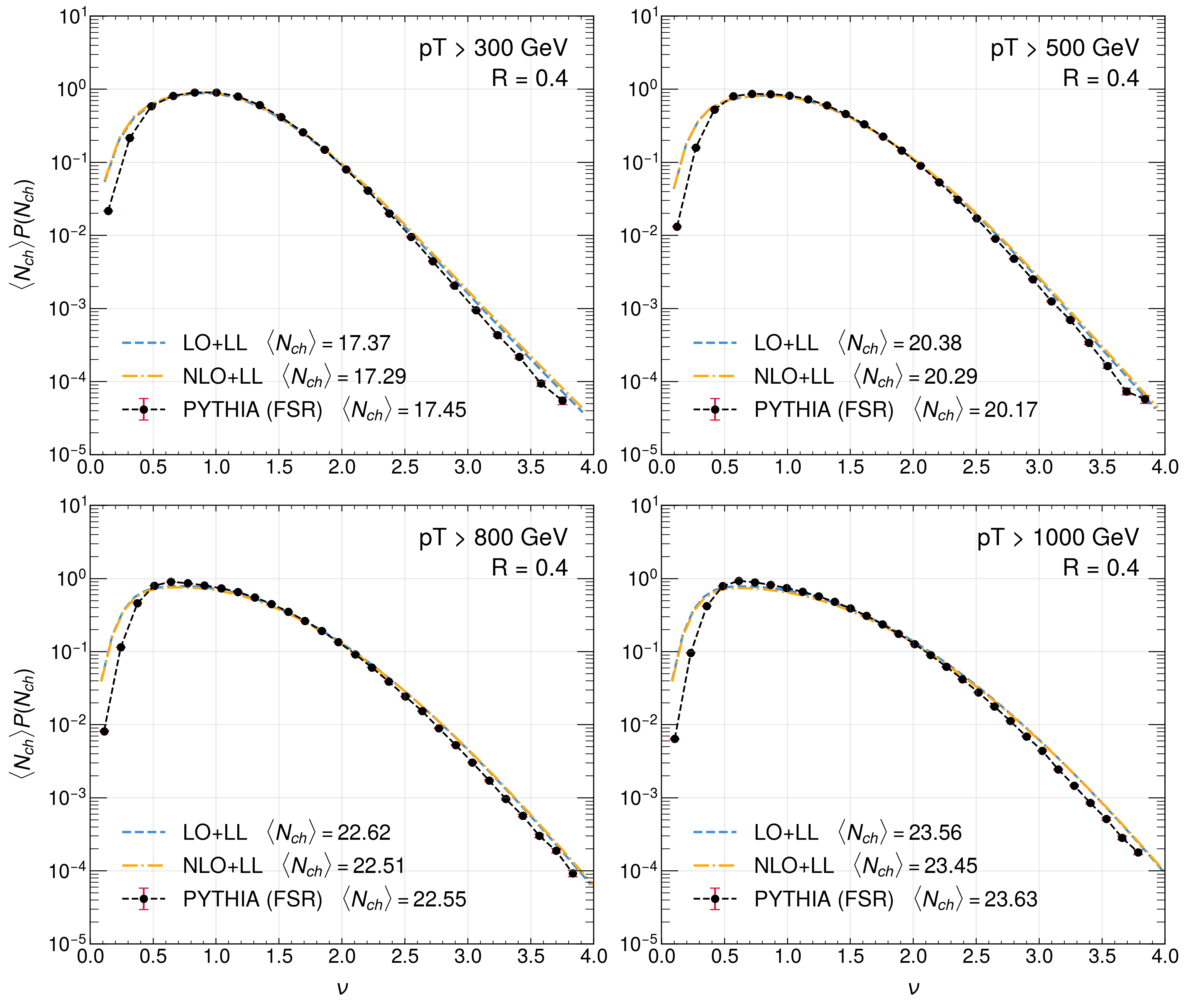}
    \caption{Rescaled charged particle multiplicity distributions $\langle N_{\rm ch}\rangle P(N_{\text{ch}})$ as a function of $\nu$ for jets with radius $R = 0.4$ with varying transverse momentum thresholds: $p_T > 300,\,500,\,800,\,1000\,\mathrm{GeV}$. Results from LO+LL (blue dashed) and NLO+LL (orange dash-dotted) are compared with \textsc{Pythia8} simulations (black markers with error bars) where only turn on FSR.}\label{fig:normo_diff_pTs}
\end{figure}
\begin{figure}
    \centering
    \includegraphics[width=1.0\textwidth]{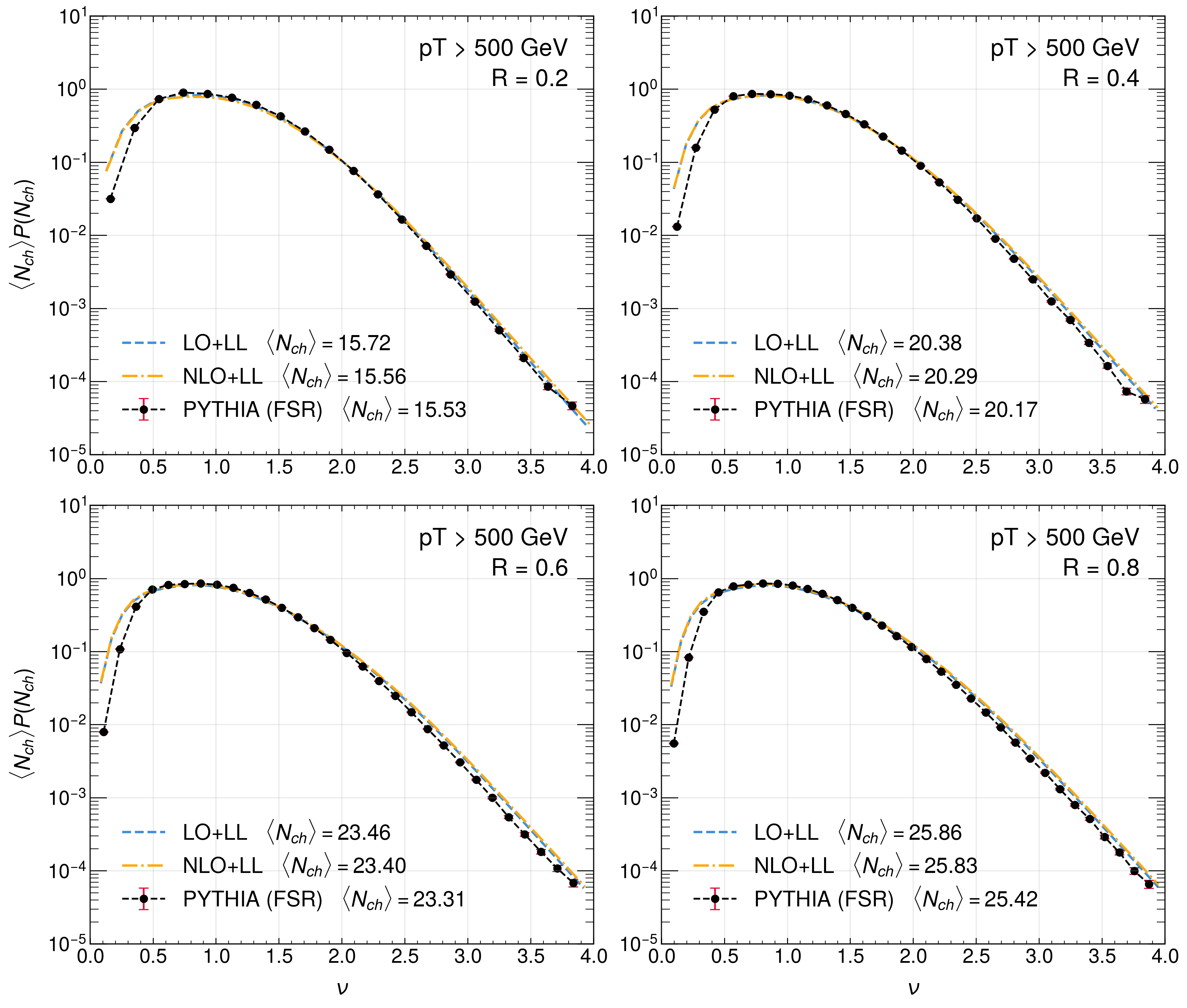}
    \caption{Rescaled charged-particle multiplicity distributions $\langle N_{\rm ch}\rangle P(N_{\text{ch}})$ as a function of $\nu$ for jets with $p_T>500$~GeV at varying jet radii $R = 0.2,0.4,0.6,0.8$. LO+LL (blue dashed) and NLO+LL (orange dash-dotted) results are compared with \textsc{Pythia8} FSR simulations (black markers).}\label{fig:normo_diff_Rs}
\end{figure}

\paragraph{Decomposition into quark and gluon jets} 
Fig.~\ref{fig:Pn_q_vs_g_R0_8} shows the decomposition of the charged-particle multiplicity distribution into contributions from quark- and gluon-initiated jets. 
For jets with $p_T > 550$~GeV at $\sqrt{s}=13$~TeV, gluon jets dominate the high-multiplicity region over quark jets by roughly an order of magnitude. 
This suggests that the resolution of the $\nu > 4$ discrepancy is likely tied to a more refined treatment of the gluon sector. 
Quark jets exhibit a smaller average multiplicity, and the inclusion of NLO corrections leads to a mild enhancement of their distribution in the high-$N_{\mathrm{ch}}$ tail, whereas the gluon-jet distribution remains essentially unchanged. 
These observations are consistent with our findings for exclusive jets in Sec.~\ref{sec:resultsExclusive}.

\paragraph{Dependence on jet $p_T$ and $R$} 
Fig.~\ref{fig:Nmean_Rs} shows the mean charged-particle multiplicity $\langle N_{\mathrm{ch}} \rangle$ as a function of jet transverse momentum for $R = 0.2, 0.4, 0.6, 0.8$, compared to \textsc{Pythia8} (FSR). As expected, $\langle N_{\mathrm{ch}} \rangle$ grows with $p_T R$, and quantitatively agrees with \textsc{Pythia8} simulations. The same set of nonperturbative initial condition is used for all cases, demonstrating the predictive power of the calculation.

Figs.~\ref{fig:normo_diff_pTs} and~\ref{fig:normo_diff_Rs} display the rescaled multiplicity distributions $\langle N_{\text{ch}} \rangle P(N_{\text{ch}})$ for $p_T > 300$, $500$, $800$, and $1000$~GeV at fixed $R=0.4$, as well as for $R = 0.2, 0.4, 0.6, 0.8$ at fixed $p_T > 500$~GeV. Across all $p_T$ and $R$, the LO+LL and NLO+LL predictions closely match the \textsc{Pythia8} results, except for the very low-multiplicity region discussed earlier. These results demonstrate that our framework consistently describes the energy and angular dependence of jet multiplicities distributions over a broad kinematic range.

\begin{figure}
    \centering
    \includegraphics[width=1.0\textwidth]{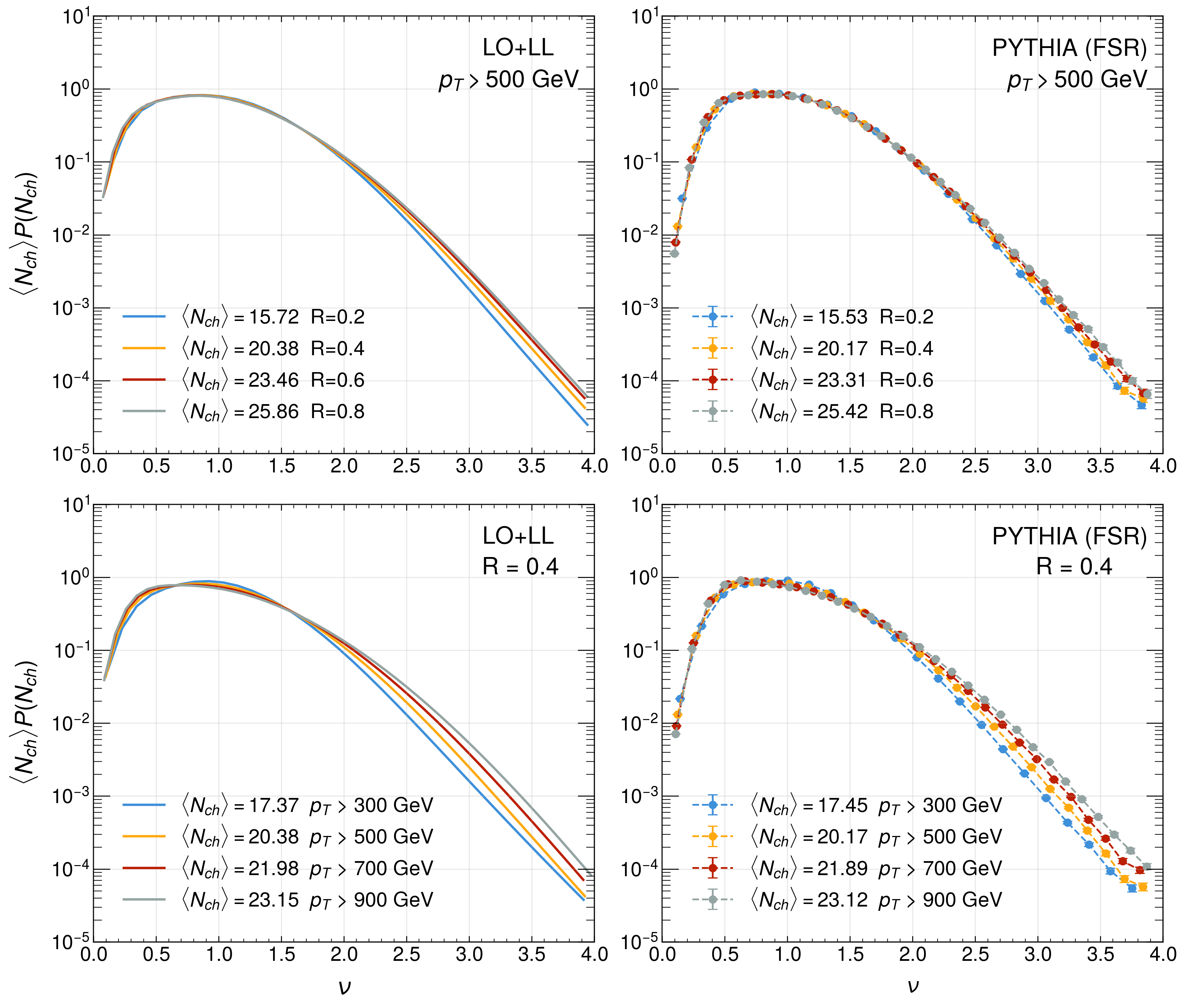}
    \caption{Rescaled charged-particle multiplicity distributions $\langle N_{\rm ch}\rangle P(N_{\text{ch}})$ as a function of $\nu$, showing only LO+LL results. The upper row corresponds to jets with $p_T > 500$ GeV and varying radii $R = 0.2, 0.4, 0.6, 0.8$, while the lower row corresponds to jets with $R = 0.4$ and varying $p_T > 300, 500, 700, 900$ GeV. Left column: LO+LL predictions. Right column: \textsc{Pythia8} simulations including only FSR.}\label{fig:inclusiveKNO}
\end{figure}

\paragraph{KNO scaling and violation} Finally, we examine Koba–Nielsen–Olesen (KNO) scaling of semi-inclusive jet multiplicities in Fig.~\ref{fig:inclusiveKNO}. 
The rescaled distributions are shown for jets with $p_T > 500$~GeV at various radii ($R = 0.2, 0.4, 0.6, 0.8$), as well as for a fixed $R = 0.4$ at $p_T = 300$, $500$, $700$, and $900$~GeV, compared with \textsc{Pythia8} (FSR) simulations. 
In both cases, KNO scaling holds approximately for $\nu < 2$, while mild violations appear for $\nu > 2$, where the distributions broaden slightly with increasing $R$ or $p_T$. 

Because gluon jets dominate the large-multiplicity tail, the primary source of the observed KNO violation at $\nu > 2$ originates from the same mechanism identified in the exclusive jet case discussed in Sec.~\ref{sec:resultsExclusive}, where the nonlinear branching equation introduces deviations from the exact scaling predicted by the double-logarithmic approximation. 
A subleading cause of KNO breaking in semi-inclusive jet samples arises from the varying mixture of quark- and gluon-initiated jets. 
Since quark and gluon jets have different mean multiplicities and distribution shapes, changes in their relative fractions with $R$ and $p_T$ naturally induce additional breaking of KNO scaling. 
Nevertheless, because the quark contribution is only about one tenth of the gluon contribution at large $\nu$, and the variation of the quark–gluon fraction with kinematics is modest, we regard this effect as subleading. 
We also observe that \textsc{Pythia8} simulations exhibit a similar systematic violation of KNO scaling, confirming that our framework captures the perturbative origin of these effects as seen in realistic parton-shower simulations.

\section{Conclusion}\label{sec:conclusion}

In this work, we have established a comprehensive theoretical framework for calculating the internal charged-particle multiplicity distribution of jets, a fundamental yet complicated QCD observable. The primary challenge lies in the interplay between non-linear perturbative parton evolution and nonperturbative hadronization in shaping the multiplicity distribution. Our approach successfully unifies these elements within a single, calculable formalism based on Soft-Collinear Effective Theory (SCET) and the generating function technique.

The core of our framework is a two-step factorization that separates the dynamics at different scales. First, we separate the evolution of the semi-inclusive jet function --- governed by DGLAP evolution in virtuality --- from the parton branching dynamics \emph{inside} the jet. Second, within the jet, we further separate the perturbative parton branching from the non-perturbative hadronization that produces the final multiplicity distribution, handled in the generating function formalism.
Our calculation incorporates NLO fixed-order corrections to the hard partonic cross section, jet functions, and the generating functions. Jet functions are improved by the LL$_R$ resummation of the jet radius, while the generating functions are evolved using a set of coupled, nonlinear equations with angular ordering and leading-order QCD splitting functions. This angular-ordered evolution is crucial for respecting QCD coherence and resumming the large soft logarithms that dominate multiplicity fluctuations. 

Our calculation naturally explains the characteristic difference between quark and gluon jets, with gluon jets exhibiting a systematically higher mean multiplicity and a broader distribution. The gluon-to-quark mean multiplicity ratio displays a clear logarithmic dependence on the jet scale $p_T R$ and is comparable to the value used in phenomenology. Besides, we provide a quantitative, first-principles demonstration of KNO scaling violation for jet multiplicities. While approximate scaling holds for $\nu\lesssim 2$, our calculations reveal a systematic broadening of the rescaled distribution for $\nu>2$ with increasing $p_T$ or $R$. This violation, mainly driven by the nonlinear nature of the branching evolution beyond the double-logarithmic approximation, is qualitatively confirmed by \textsc{Pythia8} simulations. Finally, with a single set of nonperturbative parameterization fixed by CMS data at one kinematic point ($p_T > 550$~GeV, $R=0.8$), our model successfully predicts the mean multiplicity and distribution shape across a wide range of jet transverse momenta and radii.

The current calculation also highlights specific future improvements. The most significant discrepancy lies in the high-multiplicity tail ($\nu \gtrsim 4$), where our prediction starts to systematically underestimates the CMS data. This region is particularly compelling, as it is linked to the emergence of non-trivial jet substructure in recent CMS measurements, specifically the enhancement in the near-side ridge of the two-particle correlation function within the jet-centric frame. This tension suggests that while our framework captures the dominant perturbative dynamics, more delicate details of parton showers may be required to describe regions with extreme fluctuation, e.g., the inclusion of higher-order nonlinear effects such as the intrinsic $g \to ggg$ splittings. A dedicated investigation of these enhanced branching kernels is a primary objective for our follow-up work.

In summary, we have set up first-principles calculations of jet multiplicity distributions within the jet function framework, integrating both perturbative resummation and a physically motivated nonperturbative model. This study also establishes a robust, extensible framework for a new class of jet substructure analyses. The most immediate application is to leverage this framework for calculating other jet observables \emph{conditioned} on multiplicity, providing a more differential way to investigate multi-particle correlations as function of jet internal multiplicity. Furthermore, the formalism is naturally extensible to the complex environment of heavy-ion collisions, where the angular-ordered evolution can be augmented with medium corrections, offering a unique opportunity to probe the interplay between jet quenching, hadronization, and the search for possible collective behavior inside jets.

\section*{Acknowledgements}
The authors would like to thank Lin Chen, Hai-Tao Li, Xiaoyu Liu, Adam Takacs, Wenbin Zhao, and Anwei Zhou for helpful discussions. This work is supported in part by National Natural Science Foundation of China (NSFC) under Grant Nos. 12225503 and 1234710148, and in part by China Postdoctoral Science Foundation under Grant No. 2023M742098. Some of the calculations were performed in the Nuclear Science Computing Center at Central China Normal University ($\mathrm{NSC}^3$), Wuhan, Hubei, China.

\bibliographystyle{JHEP}
\bibliography{ref}

\end{document}